\documentclass[10pt]{article}
\usepackage{graphicx}
\usepackage{amsmath}
\usepackage{amssymb}
\usepackage{caption2}
\usepackage{setspace}
\begin{document}
\bibliographystyle{prsty}
\begin{center}
{\large {\bf \sc{  Analysis of  the hidden-charm pentaquark candidates in the $J/\psi \Xi$ mass spectrum  via the  QCD sum rules }}} \\[2mm]
Zhi-Gang Wang \footnote{E-mail: zgwang@aliyun.com.  }, Yang Liu     \\
 Department of Physics, North China Electric Power University, Baoding 071003, P. R. China
\end{center}

\begin{abstract}
In this work, we construct the color $\bar{\mathbf{3}}\bar{\mathbf{3}}\bar{\mathbf{3}}$ type local five-quark currents with the light quarks $qss$ in the flavor octet, and  study the $qssc\bar{c}$ pentaquark states via  the QCD sum rules in a comprehensive way, and we emphasize that we achieve two light-flavor octets.
We obtain the mass spectrum of the hidden-charm-doubly-strange pentaquark states with the isospin-spin-parity  $IJ^{P}=\frac{1}{2}{\frac{1}{2}}^-$, $\frac{1}{2}{\frac{3}{2}}^-$ and $\frac{1}{2}{\frac{5}{2}}^-$, which can be confronted to the experimental data in the future, especially in the process $\Xi_b^- \to P_{css}^-\phi \to J/\psi \Xi^-  \phi $. As a byproduct, we observe  that the lowest hidden-charm pentaquark states are not of the scalar-diquark-scalar-diquark-antiquark type, it is not suitable  to refer to the scalar and axialvector diquarks as the "good" and "bad" diquarks, respectively.
\end{abstract}

 PACS number: 12.39.Mk, 14.20.Lq, 12.38.Lg

Key words: Pentaquark states, QCD sum rules

\section{Introduction}
In 2015,  the  LHCb collaboration  observed  two pentaquark candidates $P_c(4380)$ and $P_c(4450)$ in the $J/\psi p$ invariant mass distribution in the $\Lambda_b^0\to J/\psi pK^- $ decays, which have the preferred spin-parity   $J^P={\frac{3}{2}}^-$ and ${\frac{5}{2}}^+$, respectively, but the assignments   $J^P={\frac{3}{2}}^+$ and ${\frac{5}{2}}^-$ cannot be excluded \cite{LHCb-4380}.

In 2019, the LHCb collaboration studied the $\Lambda_b^0\to J/\psi K^- p$ decays with a data sample with an order of magnitude larger than the previously one, and observed a new pentaquark candidate $P_c(4312)$ \cite{LHCb-Pc4312}. In addition,  the LHCb collaboration confirmed the structure $P_c(4450)$, which  consists  of two narrow overlapping peaks $P_c(4440)$ and $P_c(4457)$, but the spin and parity are not determined \cite{LHCb-Pc4312}.

In 2020, the LHCb collaboration reported an evidence of a hidden-charm pentaquark candidate $P_{cs}(4459)$ with the strangeness $S=-1$ in the $J/\psi \Lambda$ mass spectrum with  a statistical significance of  $3.1\sigma$ in the $\Xi_b^- \to J/\psi K^- \Lambda$ decays, but the spin and parity are not determined \cite{LHCb-Pcs4459-2012}.

In 2021, the LHCb collaboration observed evidences  for a new structure $P_c(4337)$ in the $J/\psi p$ and $J/\psi \bar{p}$ systems  with a significance about  3.1 to 3.7$\sigma$,  which depend on the assigned $J^P$ hypothesis \cite{LHCb-Pc4337}.

In 2022, the LHCb collaboration observed an evidence for a new structure $P_{cs}(4338)$ with the favored  spin-parity $J^P={\frac{1}{2}}^-$ in the $J/\psi \Lambda$ mass distribution in the $B^- \to J/\psi \Lambda \bar{p}$ decays \cite{LHCb-Pcs4338}.

In 2025, the Belle and Belle-II collaborations observed the $\Upsilon(1{\rm S }, 2{\rm S})$ inclusive decays to the final states $J/\psi\Lambda$, and found  an evidence of the $P_{cs}(4459)$  with a local significance of $3.3\,\sigma$ \cite{Belle-Pcs4338-Pcs4459}.

Those pentaquark candidates always lie near (or just in, or slightly below) the two-particle thresholds,
\begin{eqnarray}
\bar{D}\Sigma_c(4321)&:& P_c(4312) \, ,\nonumber \\
\bar{D}\Xi_c(4337)&:& P_{cs}(4338) \, ,\nonumber \\
\bar{D}^*\Xi_c(4478)&:& P_{cs}(4459) \, ,\nonumber \\
\bar{D}\Sigma^*_c(4385)&:& P_c(4380) \, ,\nonumber \\
\bar{D}^*\Sigma_c(4462)&:& P_c(4440/4457) \, ,
\end{eqnarray}
which stimulate the possible assignments as the molecular states \cite{mole-penta-2,mole-penta-3,mole-penta-1,mole-penta-6,
mole-penta-5,mole-penta-10,
mole-penta-11,Pc4312-mole-penta-2,Pc4312-mole-penta-1,Pc4312-mole-penta-7,
Pcs4459-mole-WangZG-SR,Pcs4459-mole-CWXiao,Pcs4338-mole-FKGuo,Pcs4459-mole-FLWang,
Pc4312-mole-penta-WXW-SCPMA,Pc4312-mole-penta-WXW-IJMPA,
Pcs4338-mole-LMeng,Pcs4338-mole-XWWang}, where we present the experimental values in the brackets with the unit $\rm{MeV}$.
 As the $P_c(4337)$ is concerned, although  it lies not far away from  the  $\bar{D}^* \Lambda_c(4296)$,
  $\bar{D} \Sigma_c(4321)$ and $\bar{D} \Sigma_c^*(4385)$ thresholds, it does not lie just slightly below  any baryon-meson  threshold, it is difficult to assign it as a molecular state without introducing large coupled-channel effects to reproduce its experimental mass \cite{Coupled-Pc4337-MJYan-EPJC-2022}.

The QCD sum rules approach is a powerful theoretical tool in exploring the exotic states, such as the tetraquark states, pentaquark states, molecular states, etc \cite{WangZG-Review,WangZG-landau-PRD,Nielsen-review}.
In Refs.\cite{Pcs4459-mole-WangZG-SR,Pc4312-mole-penta-WXW-SCPMA,
Pc4312-mole-penta-WXW-IJMPA}, we adopted the scenario of hadronic molecules,
 constructed  the five-quark interpolating  currents with the definite isospins for the first time, and studied the color singlet-singlet type hidden-charm pentaquark states with the  strangeness $S=0$ and $-1$ via the QCD sum rules  comprehensively, and observed  that   except for the $P_c(4337)$, other pentaquark candidates could find their suitable positions, which is consistent with our naive expectation. In the QCD sum rules, we usually construct  the local color $\mathbf{1}\mathbf{1}$ and $\bar{\mathbf{3}}\bar{\mathbf{3}}\bar{\mathbf{3}}$ type five-quark currents, which have two color-neutral clusters and three colored clusters respectively, and couple potentially to the pentaquark states having the average spatial sizes $\sqrt{\langle r^2\rangle}$ of the same magnitudes as the conventional
 baryons \cite{WangZG-Review}. And we usually  call the color $\mathbf{1}\mathbf{1}$ type pentaquark states as  the hadronic molecular states, but they are also compact objects in the QCD sum rules, just like the $\bar{\mathbf{3}}\bar{\mathbf{3}}\bar{\mathbf{3}}$ type
pentaquark states.  All in all, at least the QCD sum rules do not support assigning the $P_c(4337)$ as the hadronic molecular state.

 While in the scenario of diquark-diquark-antiquark type (or color $\bar{\mathbf{3}}\bar{\mathbf{3}}\bar{\mathbf{3}}$ type) pentaquark states \cite{di-di-anti-penta-1,di-di-anti-penta-2,di-di-anti-penta-3,Wang1508-EPJC,
WangHuang-EPJC-1508-12,WangZG-EPJC-1509-12,WangZG-NPB-1512-32,
WangZhang-APPB,Pc4312-penta-1,Pc4312-penta-3,WZG-penta-IJMPA,
WangZG-Pcs4459-333,WangZG-Pc12-Jpsip,WangZG-Pc12-JpsiLambda} and diquark-triquark type pentaquark states \cite{di-tri-penta-1,di-tri-penta-2}, all the pentaquark (P)   states, such as the $P_c(4312)$, $P_c(4337)$, $P_{cs}(4338)$, $P_{c}(4380)$, $P_c(4440)$, $P_c(4457)$, $P_{cs}(4459)$, could find their suitable rooms to sit down \cite{WangZG-Pc12-Jpsip,WangZG-Pc12-JpsiLambda}. However, we should bear in mind the possibility that they are kinematical effects cannot be excluded, such as anomalous triangle singularities \cite{ATS-LiuXH-2016,ATS-Bayar-GuoFK-2016,ATS-LiuXH-2020}. Thereafter, we would like to use P to denote pentaquark for simplicity, and add some subscripts and superscripts if necessary.

After the observations of the hidden-charm pentaquark candidates with the strangeness $S=0$ and $-1$ in the $J/\psi p$ and $J/\psi\Lambda $ systems respectively, it is natural and interesting to
search for the hidden-charm pentaquark candidates with the strangeness $S=-2$ in the $J/\psi\Xi$ and $J/\psi \Xi^*$ systems or $S=-3$ in the $J/\psi\Omega$ system \cite{di-di-anti-penta-1,di-di-anti-penta-2,
WangZG-EPJC-1509-12,WangZG-NPB-1512-32,Pc4312-penta-1,
LiuXiang-Pcss-2021,Azizi-Pcss-2022,Oset-Pcss-2024}.

In 2025, the LHCb collaboration reported the first observation of the $\Xi_b^0 \to J/\psi \Xi^- \pi^+$ decay and the most precise measurement of the branching fraction of the $\Lambda_b^0 \to J/\psi \Xi^- K^+$ decay using proton-proton collision data corresponding to an integrated luminosity of $5.4~\rm{fb}^{-1}$ \cite{LHCb-JpsiXi-2025}.  However, they  observed no evidence for the exotic states $P_{css}$. Furthermore, the LHCb collaboration observed the
$\Lambda_b^{0} \to \Lambda_{c}^{+}D_{s}^{-}K^{+}K^{-}$ decay for the first time  and found no evidence of the  pentaquark candidates $P_{cs}(4338/4459)$  in the $\Lambda_{c}^{+}D_{s}^{-}$ mass spectrum \cite{LHCb-Lambda-Ds}.
More experimental data and more precise theoretical calculations are still needed to shed light on the nature of those P states.

 In Refs.\cite{Wang1508-EPJC,WangHuang-EPJC-1508-12,WangZG-EPJC-1509-12,
 WangZG-NPB-1512-32,WangZhang-APPB}, we resorted to the pentaquark scenario and explored the color $\bar{\mathbf{3}}\bar{\mathbf{3}}\bar{\mathbf{3}}$ type hidden-charm pentaquark states with the spin-parity  $J^P={\frac{1}{2}}^\pm$, ${\frac{3}{2}}^\pm$, ${\frac{5}{2}}^\pm$  and strangeness   $S=0,\,-1,\,-2,\,-3$ via  the QCD sum rules  systematically by accomplishing the operator product expansion up to the vacuum condensates of dimension 10.

After the discovery of the $P_c(4312)$,  we updated the old analysis and calculated  the   vacuum condensates up to dimension $13$  consistently,  and  restudied the ground state mass spectrum of the color $\bar{\mathbf{3}}\bar{\mathbf{3}}\bar{\mathbf{3}}$ type $uudc\bar{c}$ pentaquark states without restricting the isospin $I=\frac{1}{2}$, then assigned the $P_c(4312)$, $P_c(4380)$, $P_c(4440)$ and $P_c(4457)$ in a reasonable way \cite{WZG-penta-IJMPA}. In Ref.\cite{WangZG-Pc12-Jpsip}, we  exhausted   the lowest  $\bar{\mathbf{3}}\bar{\mathbf{3}}\bar{\mathbf{3}}$ type $uudc\bar{c}$ pentaquark configurations by restricting the isospin  $I=\frac{1}{2}$, and investigated the mass spectrum  via the QCD sum rules systematically, and revisited  the  assignments of the $P_{c}$ states with the isospin-spin-parity $IJ^P=\frac{1}{2}{\frac{1}{2}}^-$, $\frac{1}{2}{\frac{3}{2}}^-$ or $\frac{1}{2}{\frac{5}{2}}^-$. As a byproduct, we observed that the lowest hidden-charm pentaquark state has a mass about $4.20\,\rm{GeV}$, which is obviously smaller than the mass of the $P_c(4312)$. We emphasize that we obtain more flat Borel windows in Refs.\cite{WZG-penta-IJMPA,WangZG-Pc12-Jpsip} as the higher dimensional vacuum condensates play an important role in determining the Borel windows, for example, see Fig.\ref{mass-D10-fig} in Sect.3.

 After the discovery of the $P_{cs}(4459)$, we examined the possibility of assigning it as the isospin cousin of the $P_{c}(4312)$ by taking account of the light-flavor $SU(3)$ breaking effects \cite{WangZG-Pcs4459-333}. In Ref.\cite{WangZG-Pc12-JpsiLambda},  we exhausted   the lowest  $\bar{\mathbf{3}}\bar{\mathbf{3}}\bar{\mathbf{3}}$ type $udsc\bar{c}$ configurations with the isospin-spin-parity $IJ^P=0{\frac{1}{2}}^-$, $0{\frac{3}{2}}^-$ and $0{\frac{5}{2}}^-$ comprehensively,  and made possible assignments of the $P_{cs}(4338)$ and $P_{cs}(4459)$ consistently. We observed that the lowest state has a mass about $4.33\,\rm{GeV}$, which happens  to coincide with the mass of the $P_{cs}(4338)$.

The two-body strong decays $P_c \to J/\psi p$ and $P_{cs}\to J/\psi\Lambda$ are expected to conserve isospins  in most cases,  the $P_c$ states have the isospin $I=\frac{1}{2}$ while the $P_{cs}$ states have the isospin $I=0$.  Naively, we expect that the two-body strong decays $P_{css}\to J/\psi \Xi$ could take place naturally through the Okubo-Zweig-Iizuka super-allowed fall-apart mechanism, and would like to investigate the lowest  $\bar{\mathbf{3}}\bar{\mathbf{3}}\bar{\mathbf{3}}$ type $qssc\bar{c}$ configurations with the isospin-spin-parity $IJ^P=\frac{1}{2}{\frac{1}{2}}^-$, $\frac{1}{2}{\frac{3}{2}}^-$ and $\frac{1}{2}{\frac{5}{2}}^-$ comprehensively via the QCD sum rules, as the observations  of their isospin cousins are of crucial importance.

 The article is arranged as follows:  we obtain the QCD sum rules for the masses and pole residues of  the hidden-charm-doubly-strange pentaquark states in Sect.2;  in Sect.3, we present the numerical results and discussions; and Sect.4 is reserved for our
conclusion.

\section{QCD sum rules for  the  $qssc\bar{c}$ pentaquark states}
At the beginning point, we write down  the two-point correlation functions $\Pi(p)$, $\Pi_{\mu\nu}(p)$ and $\Pi_{\mu\nu\alpha\beta}(p)$,
\begin{eqnarray}\label{CF-Pi-Pi-Pi}
\Pi(p)&=&i\int d^4x e^{ip \cdot x} \langle0|T\left\{J(x)\bar{J}(0)\right\}|0\rangle \, ,\nonumber\\
\Pi_{\mu\nu}(p)&=&i\int d^4x e^{ip \cdot x} \langle0|T\left\{J_{\mu}(x)\bar{J}_{\nu}(0)\right\}|0\rangle \, ,\nonumber\\
\Pi_{\mu\nu\alpha\beta}(p)&=&i\int d^4x e^{ip \cdot x} \langle0|T\left\{J_{\mu\nu}(x)\bar{J}_{\alpha\beta}(0)\right\}|0\rangle \, ,
\end{eqnarray}
where the interpolating currents,
 \begin{eqnarray}
 J(x)&=&J^1(x)\, , \, J^2(x)\, , \, J^3(x)\, , \, J^4(x)\, , \, J^5(x)\, , \, J^6(x)\, , \nonumber\\
 J_\mu(x)&=&J_\mu^1(x)\, , \, J_\mu^2(x)\, , \, J_\mu^3(x)\, , \, J_\mu^4(x)\, , \, J_\mu^5(x)\, , \, J_\mu^6(x)\, , \, J_\mu^7(x)\, ,  \nonumber\\
 J_{\mu\nu}(x)&=&J_{\mu\nu}^1(x)\, , \, J_{\mu\nu}^2(x)\, ,\, J_{\mu\nu}^3(x)\, ,\, J_{\mu\nu}^4(x)\, ,\, J_{\mu\nu}^5(x)\, ,
 \end{eqnarray}
 with
\begin{eqnarray}\label{Current-12}
J^1(x)&=&\varepsilon^{ila} \varepsilon^{ijk}\varepsilon^{lmn}  q^T_j(x) C\gamma_5 s_k(x)\,s^T_m(x) C\gamma_5 c_n(x)\, C\bar{c}^{T}_{a}(x) \, , \nonumber \\
J^2(x)&=&\varepsilon^{ila} \varepsilon^{ijk}\varepsilon^{lmn}  q^T_j(x) C\gamma_5 s_k(x)\,s^T_m(x) C\gamma_\mu c_n(x)\,\gamma_5 \gamma^\mu C\bar{c}^{T}_{a}(x) \, ,\nonumber \\
J^{3}(x)&=&\frac{\varepsilon^{ila} \varepsilon^{ijk}\varepsilon^{lmn}}{\sqrt{2}} \left[ s^T_j(x) C\gamma_\mu s_k(x)q^T_m(x) C\gamma^\mu c_n(x)-s^T_j(x) C\gamma_\mu q_k(x)s^T_m(x) C\gamma^\mu c_n(x) \right]  C\bar{c}^{T}_{a}(x) \, , \nonumber\\
J^{4}(x)&=&\frac{\varepsilon^{ila} \varepsilon^{ijk}\varepsilon^{lmn}}{\sqrt{2}} \left[ s^T_j(x) C\gamma_\mu s_k(x) q^T_m(x) C\gamma_5 c_n(x)-s^T_j(x) C\gamma_\mu q_k(x) s^T_m(x) C\gamma_5 c_n(x)\right] \gamma_5 \gamma^\mu  C\bar{c}^{T}_{a}(x) \, ,  \nonumber\\
 J^{5}(x)&=&\varepsilon^{ila} \varepsilon^{ijk}\varepsilon^{lmn}  s^T_j(x) C\gamma_\mu s_k(x)\, q^T_m(x) C\gamma^\mu c_n(x) C\bar{c}^{T}_{a}(x) \, , \nonumber\\
J^{6}(x)&=&\varepsilon^{ila} \varepsilon^{ijk}\varepsilon^{lmn}  s^T_j(x) C\gamma_\mu s_k(x) \,q^T_m(x) C\gamma_5 c_n(x)  \gamma_5 \gamma^\mu C\bar{c}^{T}_{a}(x) \, ,
\end{eqnarray}
 for the isospin-spin $(I,J)=(\frac{1}{2},\frac{1}{2})$,
\begin{eqnarray}\label{Current-32}
 J^1_{\mu}(x)&=&\varepsilon^{ila} \varepsilon^{ijk}\varepsilon^{lmn}
 q^T_j(x) C\gamma_5s_k(x)\,s^T_m(x) C\gamma_\mu c_n(x)\, C\bar{c}^{T}_{a}(x) \, , \nonumber \\
J^{2}_{\mu}(x)&=&\frac{\varepsilon^{ila} \varepsilon^{ijk}\varepsilon^{lmn}}{\sqrt{2}} \left[ s^T_j(x) C\gamma_\mu s_k(x) q^T_m(x) C\gamma_5 c_n(x) -s^T_j(x) C\gamma_\mu q_k(x) s^T_m(x) C\gamma_5 c_n(x)\right]   C\bar{c}^{T}_{a}(x) \, , \nonumber \\
 J^{3}_{\mu}(x)&=&\frac{\varepsilon^{ila} \varepsilon^{ijk}\varepsilon^{lmn}}{\sqrt{2}} \left[ s^T_j(x) C\gamma_\mu s_k(x)q^T_m(x) C\gamma_\alpha c_n(x)-s^T_j(x) C\gamma_\mu q_k(x)s^T_m(x) C\gamma_\alpha c_n(x) \right] \gamma_5\gamma^\alpha C\bar{c}^{T}_{a}(x) \, , \nonumber\\
J^{4}_{\mu}(x)&=&\frac{\varepsilon^{ila} \varepsilon^{ijk}\varepsilon^{lmn}}{\sqrt{2}} \left[ s^T_j(x) C\gamma_\alpha s_k(x)q^T_m(x) C\gamma_\mu c_n(x)-s^T_j(x) C\gamma_\alpha q_k(x)s^T_m(x) C\gamma_\mu c_n(x) \right] \gamma_5\gamma^\alpha C\bar{c}^{T}_{a}(x) \, ,\nonumber\\
 J^{5}_{\mu}(x)&=&\varepsilon^{ila} \varepsilon^{ijk}\varepsilon^{lmn} \ s^T_j(x) C\gamma_\mu s_k(x)\, q^T_m(x) C\gamma_5 c_n(x)    C\bar{c}^{T}_{a}(x) \, , \nonumber \\
J^{6}_{\mu}(x)&=&\varepsilon^{ila} \varepsilon^{ijk}\varepsilon^{lmn}  s^T_j(x) C\gamma_\mu s_k(x)\, q^T_m(x) C\gamma_\alpha c_n(x)\gamma_5\gamma^\alpha C\bar{c}^{T}_{a}(x) \, , \nonumber\\
J^{7}_{\mu}(x)&=&\varepsilon^{ila} \varepsilon^{ijk}\varepsilon^{lmn}  s^T_j(x) C\gamma_\alpha s_k(x)\, q^T_m(x) C\gamma_\mu c_n(x) \gamma_5\gamma^\alpha C\bar{c}^{T}_{a}(x) \, ,
\end{eqnarray}
 for the isospin-spin $(I,J)=(\frac{1}{2},\frac{3}{2})$,
\begin{eqnarray} \label{Current-52}
J^1_{\mu\nu}(x)&=&\frac{\varepsilon^{ila} \varepsilon^{ijk}\varepsilon^{lmn} }{\sqrt{2}} q^T_j(x) C\gamma_5  s_k(x) \, s^T_m(x) C\gamma_\mu c_n(x)\, \gamma_5\gamma_{\nu}C\bar{c}^{T}_{a}(x)+(\mu \leftrightarrow \nu) \, ,\nonumber\\
J^2_{\mu\nu}(x)&=&\frac{\varepsilon^{ila} \varepsilon^{ijk}\varepsilon^{lmn} }{2}\, s^T_j(x) C\gamma_\mu s_k(x)\, q^T_m(x) C\gamma_5 c_n(x)\, \gamma_5\gamma_{\nu}C\bar{c}^{T}_{a}(x)+(\mu \leftrightarrow \nu) \, ,\nonumber\\
&&-\frac{\varepsilon^{ila} \varepsilon^{ijk}\varepsilon^{lmn} }{2}\, s^T_j(x) C\gamma_\mu q_k(x)\, s^T_m(x) C\gamma_5 c_n(x)\, \gamma_5\gamma_{\nu}C\bar{c}^{T}_{a}(x)+(\mu \leftrightarrow \nu) \, ,\nonumber\\
J^3_{\mu\nu}(x)&=&\frac{\varepsilon^{ila} \varepsilon^{ijk}\varepsilon^{lmn}}{2} s^T_j(x) C\gamma_\mu s_k(x)\, q^T_m(x) C\gamma_\nu c_n(x)  C\bar{c}^{T}_{a}(x)+(\mu \leftrightarrow \nu)\, ,\nonumber\\
&&-\frac{\varepsilon^{ila} \varepsilon^{ijk}\varepsilon^{lmn}}{2} s^T_j(x) C\gamma_\mu q_k(x)\, s^T_m(x) C\gamma_\nu c_n(x)  C\bar{c}^{T}_{a}(x)+(\mu \leftrightarrow \nu)\, ,\nonumber\\
J^4_{\mu\nu}(x)&=&\frac{\varepsilon^{ila} \varepsilon^{ijk}\varepsilon^{lmn} }{\sqrt{2}}\, q^T_j(x) C\gamma_\mu s_k(x)\, s^T_m(x) C\gamma_5 c_n(x)\, \gamma_5\gamma_{\nu}C\bar{c}^{T}_{a}(x)+(\mu \leftrightarrow \nu) \, ,\nonumber\\
J^5_{\mu\nu}(x)&=&\frac{\varepsilon^{ila} \varepsilon^{ijk}\varepsilon^{lmn}}{\sqrt{2}} s^T_j(x) C\gamma_\mu s_k(x)\, q^T_m(x) C\gamma_\nu c_n(x)  C\bar{c}^{T}_{a}(x)+(\mu \leftrightarrow \nu)\, ,
\end{eqnarray}
for the isospin-spin $(I,J)=(\frac{1}{2},\frac{5}{2})$,
$q=u$, $d$, the $i$, $j$, $k$, $l$, $m$, $n$ and $a$ are color indices, the $C$ is the charge conjugation matrix.

For the five-quark systems with the symbolic valence quark structure $qqqQ\bar{Q}$, where $q=u$, $d$ or $s$, $Q=b$ or $c$, the components $qqq$ have the light-flavor $SU(3)$ symmetry while the components $Q\bar{Q}$ obey the heavy quark symmetry. Thus,
\begin{eqnarray}\label{two-octet}
{\mathbf{3}}\otimes {\mathbf{3}}\otimes {\mathbf{3}} &\to & \left({\bar{\mathbf{3}}} \oplus {\mathbf{6}} \right) \otimes {\mathbf{3}}\, , \nonumber\\
&\to& \left({\mathbf{1}} \oplus {\mathbf{8}}_1 \right)\oplus \left({\mathbf{8}}_2\oplus {\mathbf{10}}\right)\, ,
\end{eqnarray}
there are two octets, which could mix with each other. Generally speaking,  the light baryons only have one octet, however, the $\Lambda(1405)$ maybe have two structures \cite{Messiner-two-octet}, such a possibility cannot be excluded. The currents $J^{1/2/3/4}(x)$, $J^{1/2/3/4}_\mu(x)$ and $J^{1/2/3}_{\mu\nu}(x)$ belong to the octet ${\mathbf{8}}_1$, analogous currents can be found in Refs.\cite{WangZG-Pc12-Jpsip,WangZG-Pc12-JpsiLambda}, while the currents $J^{5/6}(x)$, $J^{5/6/7}_\mu(x)$ and $J^{4/5}_{\mu\nu}(x)$ belong to the octet ${\mathbf{8}}_2$, analogous currents can be constructed routinely. With the simple replacements,
\begin{eqnarray}
s \to q\, , \, q\to s \, ,
\end{eqnarray}
we obtain the corresponding currents with the symbolic quark structure  $qqs c\bar{c}$, and the predictions would be presented elsewhere \cite{WangZG-Pc12-JpsiSigma}. The calculations are straightforward but highly non-trivial in studying the light-flavor $SU(3)$ mass-breaking effects.    For the currents in the decuplet representation, one is suggested to refer to  Refs.\cite{WangZG-Review,WangZG-EPJC-1509-12,WangZG-NPB-1512-32,WZG-penta-IJMPA}.

If we take  the $S_L$ and $S_H$
to represent  the spins of the light  and heavy diquarks respectively, the  $\varepsilon^{ijk}q^T_jC\gamma_{5}s_k$, $\varepsilon^{ijk}q^T_jC\gamma_{\mu}s_k$ and  $\varepsilon^{ijk}s^T_jC\gamma_{\mu}s_k$ have the spins $S_L=0$, $1$ and $1$, respectively, the $\varepsilon^{ijk}q^T_jC\gamma_5c_k$, $\varepsilon^{ijk}s^T_jC\gamma_5c_k$, $\varepsilon^{ijk}q^T_jC\gamma_{\mu}c_k$ and   $\varepsilon^{ijk}s^T_jC\gamma_{\mu}c_k$ have the spins $S_H=0$, $0$, $1$ and $1$, respectively. Here the spins $S_{L/H}=0$ and $1$ correspond to the scalar (S) and  axialvector (A) diquarks, respectively.
A light and  a heavy diquark form a tetraquark in the color $\mathbf{3}$ with  angular momentum $\vec{J}_{LH}=\vec{S}_L+\vec{S}_H$ with  the values $J_{LH}=0$, $1$ or $2$.
The anti-charm quark operators $C\bar{c}_a^T$ and $\gamma_5\gamma_{\mu}C\bar{c}_a^T$ have the spin-parity $J^P={\frac{1}{2}}^-$ and ${\frac{3}{2}}^-$, respectively. Therefore the total angular momentums   $\vec{J}=\vec{J}_{LH}+\vec{J}_{\bar{c}}$ have the values $J=\frac{1}{2}$, $\frac{3}{2}$ or $\frac{5}{2}$, which are shown explicitly in Table \ref{current-pentaQ}. Even if we only take the most stable diquark configurations, i.e. $S$ and $A$ diquarks, we obtain copious spectroscopy for the hidden-charm pentaquark states due to their complex structures, all the observed $P_c$ and $P_{cs}$ states could find their suitable rooms to sit down \cite{WangZG-Pc12-Jpsip,WangZG-Pc12-JpsiLambda}. In the present work, we study the ground states, and set the orbital angular momentum  to be zero. There might be explicit (or implicit) P-waves in the diquarks, between two diquarks, between the diquark and antiquark, the additional P-waves would lead to larger pentaquark masses.

\begin{table}
\begin{center}
\begin{tabular}{|c|c|c|c|c|c|c|c|c|}\hline\hline
$[qq][qc]\bar{c}$ ($S_L$, $S_H$, $J_{LH}$, $J$)  & $J^{P}$             & Currents              \\ \hline

$[qs][sc]\bar{c}$ ($0$, $0$, $0$, $\frac{1}{2}$) &${\frac{1}{2}}^{-}$  &$J^1(x)$     \\

$[qs][sc]\bar{c}$ ($0$, $1$, $1$, $\frac{1}{2}$) &${\frac{1}{2}}^{-}$  &$J^2(x)$    \\

$[ss][qc]\bar{c}-[sq][sc]\bar{c}$ ($1$, $1$, $0$, $\frac{1}{2}$) &${\frac{1}{2}}^{-}$  &$J^3(x)$        \\

$[ss][qc]\bar{c}-[sq][sc]\bar{c}$ ($1$, $0$, $1$, $\frac{1}{2}$) &${\frac{1}{2}}^{-}$  &$J^4(x)$             \\

$[ss][qc]\bar{c}$ ($1$, $1$, $0$, $\frac{1}{2}$) &${\frac{1}{2}}^{-}$  &$J^5(x)$        \\

$[ss][qc]\bar{c}$ ($1$, $0$, $1$, $\frac{1}{2}$) &${\frac{1}{2}}^{-}$  &$J^6(x)$             \\ \hline

$[qs][sc]\bar{c}$ ($0$, $1$, $1$, $\frac{3}{2}$) &${\frac{3}{2}}^{-}$ &$J^1_\mu(x)$  \\

$[ss][qc]\bar{c}-[sq][sc]\bar{c}$ ($1$, $0$, $1$, $\frac{3}{2}$) &${\frac{3}{2}}^{-}$ &$J^2_\mu(x)$          \\

$[ss][qc]\bar{c}-[sq][sc]\bar{c}$ ($1$, $1$, $2$, $\frac{3}{2}$)${}_3$ &${\frac{3}{2}}^{-}$  &$J^3_\mu(x)$   \\

$[ss][qc]\bar{c}-[sq][sc]\bar{c}$ ($1$, $1$, $2$, $\frac{3}{2}$)${}_4$ &${\frac{3}{2}}^{-}$  &$J^4_\mu(x)$   \\

$[ss][qc]\bar{c}$ ($1$, $0$, $1$, $\frac{3}{2}$) &${\frac{3}{2}}^{-}$ &$J^5_\mu(x)$          \\

$[ss][qc]\bar{c}$ ($1$, $1$, $2$, $\frac{3}{2}$)${}_6$ &${\frac{3}{2}}^{-}$  &$J^6_\mu(x)$   \\

$[ss][qc]\bar{c}$ ($1$, $1$, $2$, $\frac{3}{2}$)${}_7$ &${\frac{3}{2}}^{-}$  &$J^7_\mu(x)$   \\ \hline

$[qs][sc]\bar{c}$ ($0$, $1$, $1$, $\frac{5}{2}$) &${\frac{5}{2}}^{-}$  &$J^1_{\mu\nu}(x)$     \\

$[ss][qc]\bar{c}-[sq][sc]\bar{c}$ ($1$, $0$, $1$, $\frac{5}{2}$) &${\frac{5}{2}}^{-}$  &$J^2_{\mu\nu}(x)$    \\

$[ss][qc]\bar{c}-[sq][sc]\bar{c}$ ($1$, $1$, $2$, $\frac{5}{2}$) &${\frac{5}{2}}^{-}$  &$J^3_{\mu\nu}(x)$   \\

$[qs][sc]\bar{c}$ ($1$, $0$, $1$, $\frac{5}{2}$) &${\frac{5}{2}}^{-}$  &$J^4_{\mu\nu}(x)$    \\

$[ss][qc]\bar{c}$ ($1$, $1$, $2$, $\frac{5}{2}$) &${\frac{5}{2}}^{-}$  &$J^5_{\mu\nu}(x)$   \\
\hline\hline
\end{tabular}
\end{center}
\caption{ The valence quark structures and spin-parity of the interpolating  currents.  }\label{current-pentaQ}
\end{table}

The currents $J(x)$, $J_\mu(x)$ and $J_{\mu\nu}(x)$ have the spin-parity
$J^P={\frac{1}{2}}^-$, ${\frac{3}{2}}^-$ and ${\frac{5}{2}}^-$, respectively, and
are expected to couple potentially to the hidden-charm-doubly-strange  pentaquark states with negative  and positive parity \cite{WangZG-Review,Wang1508-EPJC},
\begin{eqnarray}\label{Coupling12}
\langle 0| J (0)|P_{\frac{1}{2}}^{-}(p)\rangle &=&\lambda^{-}_{\frac{1}{2}} U^{-}(p,s) \, , \nonumber \\
\langle 0| J (0)|P_{\frac{1}{2}}^{+}(p)\rangle &=&\lambda^{+}_{\frac{1}{2}} i\gamma_5 U^{+}(p,s) \, ,
\end{eqnarray}
\begin{eqnarray}
\langle 0| J_{\mu} (0)|P_{\frac{3}{2}}^{-}(p)\rangle &=&\lambda^{-}_{\frac{3}{2}} U^{-}_\mu(p,s) \, ,  \nonumber \\
\langle 0| J_{\mu} (0)|P_{\frac{3}{2}}^{+}(p)\rangle &=&\lambda^{+}_{\frac{3}{2}}i\gamma_5 U^{+}_\mu(p,s) \, ,  \nonumber \\
\langle 0| J_{\mu} (0)|P_{\frac{1}{2}}^{+}(p)\rangle &=&f^{+}_{\frac{1}{2}}p_\mu U^{+}(p,s) \, , \nonumber \\
\langle 0| J_{\mu} (0)|P_{\frac{1}{2}}^{-}(p)\rangle &=&f^{-}_{\frac{1}{2}}p_\mu i\gamma_5 U^{-}(p,s) \, ,
\end{eqnarray}
\begin{eqnarray}\label{Coupling52}
\langle 0| J_{\mu\nu} (0)|P_{\frac{5}{2}}^{-}(p)\rangle &=&\sqrt{2}\lambda^{-}_{\frac{5}{2}} U^{-}_{\mu\nu}(p,s) \, ,\nonumber\\
\langle 0| J_{\mu\nu} (0)|P_{\frac{5}{2}}^{+}(p)\rangle &=&\sqrt{2}\lambda^{+}_{\frac{5}{2}}i\gamma_5 U^{+}_{\mu\nu}(p,s) \, ,\nonumber\\
\langle 0| J_{\mu\nu} (0)|P_{\frac{3}{2}}^{+}(p)\rangle &=&f^{+}_{\frac{3}{2}} \left[p_\mu U^{+}_{\nu}(p,s)+p_\nu U^{+}_{\mu}(p,s)\right] \, , \nonumber\\
\langle 0| J_{\mu\nu} (0)|P_{\frac{3}{2}}^{-}(p)\rangle &=&f^{-}_{\frac{3}{2}}i\gamma_5 \left[p_\mu U^{-}_{\nu}(p,s)+p_\nu U^{-}_{\mu}(p,s)\right] \, , \nonumber\\
\langle 0| J_{\mu\nu} (0)|P_{\frac{1}{2}}^{-}(p)\rangle &=&g^{-}_{\frac{1}{2}}p_\mu p_\nu U^{-}(p,s) \, , \nonumber\\
\langle 0| J_{\mu\nu} (0)|P_{\frac{1}{2}}^{+}(p)\rangle &=&g^{+}_{\frac{1}{2}}p_\mu p_\nu i\gamma_5 U^{+}(p,s) \, ,
\end{eqnarray}
where  the superscripts $\pm$  denote  the  parities, the subscripts $\frac{1}{2}$, $\frac{3}{2}$ and $\frac{5}{2}$  denote  the spins,  the $\lambda$, $f$ and $g$ are the pole residues,  because   multiplying $i \gamma_{5}$ to the interpolating currents   $J(x)$, $J_\mu(x)$ and $J_{\mu\nu}(x)$ changes their parities. The $U^\pm(p,s)$,  $U^{\pm}_\mu(p,s)$ and $U^{\pm}_{\mu\nu}(p,s)$ are Dirac and Rarita-Schwinger spinors respectively \cite{WangZG-Review,Wang1508-EPJC}.

On the hadron  side, we insert  a complete set  of intermediate
hidden-charm-doubly-strange  pentaquark states with the same quantum numbers as the currents  $J(x)$, $i\gamma_5 J(x)$, $J_{\mu}(x)$, $i\gamma_5 J_{\mu}(x)$, $J_{\mu\nu}(x)$  and $i\gamma_5 J_{\mu\nu}(x)$ into the correlation functions
$\Pi(p)$, $\Pi_{\mu\nu}(p)$ and $\Pi_{\mu\nu\alpha\beta}(p)$ to obtain the hadronic representation
\cite{SVZ79-1,SVZ79-2,PRT85},  isolate the  lowest  states, and obtain the results:
\begin{eqnarray}\label{CF-Hadron-12}
\Pi(p) & = & {\lambda^{-}_{\frac{1}{2}}}^2  {\!\not\!{p}+ M_{-} \over M_{-}^{2}-p^{2}  }+  {\lambda^{+}_{\frac{1}{2}}}^2  {\!\not\!{p}- M_{+} \over M_{+}^{2}-p^{2}  } +\cdots  \, ,\nonumber\\
&=&\Pi_{\frac{1}{2}}^1(p^2)\!\not\!{p}+\Pi_{\frac{1}{2}}^0(p^2)\, ,
 \end{eqnarray}
\begin{eqnarray}\label{CF-Hadron-32}
 \Pi_{\mu\nu}(p) & = & {\lambda^{-}_{\frac{3}{2}}}^2  {\!\not\!{p}+ M_{-} \over M_{-}^{2}-p^{2}  } \left(- g_{\mu\nu}+\frac{\gamma_\mu\gamma_\nu}{3}+\frac{2p_\mu p_\nu}{3p^2}-\frac{p_\mu\gamma_\nu-p_\nu \gamma_\mu}{3\sqrt{p^2}}
\right)\nonumber\\
&&+  {\lambda^{+}_{\frac{3}{2}}}^2  {\!\not\!{p}- M_{+} \over M_{+}^{2}-p^{2}  } \left(- g_{\mu\nu}+\frac{\gamma_\mu\gamma_\nu}{3}+\frac{2p_\mu p_\nu}{3p^2}-\frac{p_\mu\gamma_\nu-p_\nu \gamma_\mu}{3\sqrt{p^2}}
\right)   \nonumber \\
& &+ {f^{+}_{\frac{1}{2}}}^2  {\!\not\!{p}+ M_{+} \over M_{+}^{2}-p^{2}  } p_\mu p_\nu+  {f^{-}_{\frac{1}{2}}}^2  {\!\not\!{p}- M_{-} \over M_{-}^{2}-p^{2}  } p_\mu p_\nu  +\cdots  \, , \nonumber\\
&=&\left[\Pi_{\frac{3}{2}}^1(p^2)\!\not\!{p}+\Pi_{\frac{3}{2}}^0(p^2)\right]\left(- g_{\mu\nu}\right)+\cdots\, ,
\end{eqnarray}
\begin{eqnarray}\label{CF-Hadron-52}
\Pi_{\mu\nu\alpha\beta}(p) & = &2{\lambda^{-}_{\frac{5}{2}}}^2  {\!\not\!{p}+ M_{-} \over M_{-}^{2}-p^{2}  } \left[\frac{ \widetilde{g}_{\mu\alpha}\widetilde{g}_{\nu\beta}+\widetilde{g}_{\mu\beta}\widetilde{g}_{\nu\alpha}}{2}-\frac{\widetilde{g}_{\mu\nu}\widetilde{g}_{\alpha\beta}}{5}-\frac{1}{10}\left( \gamma_{\mu}\gamma_{\alpha}+\frac{\gamma_{\mu}p_{\alpha}-\gamma_{\alpha}p_{\mu}}{\sqrt{p^2}}-\frac{p_{\mu}p_{\alpha}}{p^2}\right)\widetilde{g}_{\nu\beta}\right.\nonumber\\
&&\left.-\frac{1}{10}\left( \gamma_{\nu}\gamma_{\alpha}+\frac{\gamma_{\nu}p_{\alpha}-\gamma_{\alpha}p_{\nu}}{\sqrt{p^2}}-\frac{p_{\nu}p_{\alpha}}{p^2}\right)\widetilde{g}_{\mu\beta}
+\cdots\right]\nonumber\\
&&+  2 {\lambda^{+}_{\frac{5}{2}}}^2  {\!\not\!{p}- M_{+} \over M_{+}^{2}-p^{2}  } \left[\frac{ \widetilde{g}_{\mu\alpha}\widetilde{g}_{\nu\beta}+\widetilde{g}_{\mu\beta}\widetilde{g}_{\nu\alpha}}{2}
-\frac{\widetilde{g}_{\mu\nu}\widetilde{g}_{\alpha\beta}}{5}-\frac{1}{10}\left( \gamma_{\mu}\gamma_{\alpha}+\frac{\gamma_{\mu}p_{\alpha}-\gamma_{\alpha}p_{\mu}}{\sqrt{p^2}}-\frac{p_{\mu}p_{\alpha}}{p^2}\right)\widetilde{g}_{\nu\beta}\right.\nonumber\\
&&\left.
-\frac{1}{10}\left( \gamma_{\nu}\gamma_{\alpha}+\frac{\gamma_{\nu}p_{\alpha}-\gamma_{\alpha}p_{\nu}}{\sqrt{p^2}}-\frac{p_{\nu}p_{\alpha}}{p^2}\right)\widetilde{g}_{\mu\beta}
 +\cdots\right]   \nonumber\\
 && +{f^{+}_{\frac{3}{2}}}^2  {\!\not\!{p}+ M_{+} \over M_{+}^{2}-p^{2}  } \left[ p_\mu p_\alpha \left(- g_{\nu\beta}+\frac{\gamma_\nu\gamma_\beta}{3}+\frac{2p_\nu p_\beta}{3p^2}-\frac{p_\nu\gamma_\beta-p_\beta \gamma_\nu}{3\sqrt{p^2}}
\right)+\cdots \right]\nonumber\\
&&+  {f^{-}_{\frac{3}{2}}}^2  {\!\not\!{p}- M_{-} \over M_{-}^{2}-p^{2}  } \left[ p_\mu p_\alpha \left(- g_{\nu\beta}+\frac{\gamma_\nu\gamma_\beta}{3}+\frac{2p_\nu p_\beta}{3p^2}-\frac{p_\nu\gamma_\beta-p_\beta \gamma_\nu}{3\sqrt{p^2}}
\right)+\cdots \right]   \nonumber \\
& &+ {g^{-}_{\frac{1}{2}}}^2  {\!\not\!{p}+ M_{-} \over M_{-}^{2}-p^{2}  } p_\mu p_\nu p_\alpha p_\beta+  {g^{+}_{\frac{1}{2}}}^2  {\!\not\!{p}- M_{+} \over M_{+}^{2}-p^{2}  } p_\mu p_\nu p_\alpha p_\beta  +\cdots \, , \nonumber\\
& = & \left[\Pi_{\frac{5}{2}}^1(p^2)\!\not\!{p}+\Pi_{\frac{5}{2}}^0(p^2)\right]\left( g_{\mu\alpha}g_{\nu\beta}+g_{\mu\beta}g_{\nu\alpha}\right)  +\cdots \, ,
 \end{eqnarray}
where $\widetilde{g}_{\mu\nu}=g_{\mu\nu}-\frac{p_{\mu}p_{\nu}}{p^2}$, the $M_{\mp}$ denote the masses of the hidden-charm pentaquark states with the negative and positive parities, respectively. We select  the components $\Pi_{\frac{1}{2}}^1(p^2)$, $\Pi_{\frac{1}{2}}^0(p^2)$, $\Pi_{\frac{3}{2}}^1(p^2)$, $\Pi_{\frac{3}{2}}^0(p^2)$, $\Pi_{\frac{5}{2}}^1(p^2)$ and  $\Pi_{\frac{5}{2}}^0(p^2)$ to avoid possible contaminations from other hidden-charm-doubly-strange pentaquark states with different spins.

Then we obtain the hadronic spectral densities  through  dispersion relation,
\begin{eqnarray}
\frac{{\rm Im}\Pi^1_j(s)}{\pi}&=& \lambda_{-}^2 \delta\left(s-M_{-}^2\right)+\lambda_{+}^2 \delta\left(s-M_{+}^2\right) =\, \rho^1_{H}(s) \, , \\
\frac{{\rm Im}\Pi^0_j(s)}{\pi}&=&M_{-}\lambda_{-}^2 \delta\left(s-M_{-}^2\right)-M_{+}\lambda_{+}^2 \delta\left(s-M_{+}^2\right)
=\rho^0_{H}(s) \, ,
\end{eqnarray}
where $j=\frac{1}{2}$, $\frac{3}{2}$, $\frac{5}{2}$, we introduce the subscript $H$ to denote  the hadron side,
then we resort to the  weight functions $\sqrt{s}\exp\left(-\frac{s}{T^2}\right)$ and $\exp\left(-\frac{s}{T^2}\right)$ to obtain the QCD sum rules
on the hadron side,
\begin{eqnarray}
\int_{4m_c^2}^{s_0}ds \left[\sqrt{s}\,\rho^1_{H}(s)+\rho^0_{H}(s)\right]\exp\left( -\frac{s}{T^2}\right)
&=&2M_{-}\lambda_{-}^2\exp\left( -\frac{M_{-}^2}{T^2}\right) \, ,
\end{eqnarray}
\begin{eqnarray}
\int_{4m_c^2}^{s^\prime_0}ds \left[\sqrt{s}\,\rho^1_{H}(s)-\rho^0_{H}(s)\right]\exp\left( -\frac{s}{T^2}\right)
&=&2M_{+}\lambda_{+}^2\exp\left( -\frac{M_{+}^2}{T^2}\right) \, ,
\end{eqnarray}
where the $s_0$ and $s_0^\prime$ are the continuum threshold parameters for the hidden-charm-doubly-strange pentaquark states with negative and positive parity, respectively;  and the $T^2$ is the Borel parameter. The $T^2$ corresponds to the usually called Borel transformation $B_{T^2}$ with respect to the variable $P^2=-p^2$,
\begin{eqnarray}
B_{T^2} {\Pi(p^2)}&=& \lim_{\stackrel{P^2,n \to \infty}{P^2/n=T^2}} \frac{(P^2)^{(n+1)}}{\Gamma(n+1)}\left(-\frac{d}{dP^2}\right)^n \frac{1}{\pi}\int_{\Delta^2}^{\infty} ds \frac{{\rm Im}\Pi(s)}{s+P^2}\, ,\nonumber\\
&=&\frac{1}{\pi}\int_{\Delta^2}^{s_0} ds \,{\rm Im}\Pi(s)\, \exp\left(-\frac{s}{T^2} \right)+\frac{1}{\pi}\int_{s_0}^{\infty} ds \,{\rm Im}\Pi(s)\, \exp\left(-\frac{s}{T^2} \right)\, ,
\end{eqnarray}
where the $\Pi(p^2)$ can be any Lorentz invariant correlation functions, and the $\Delta^2$ is the corresponding lower thresholds.
Thus we  distinguish  the  contributions  of the hidden-charm-doubly-strange pentaquark states with negative and positive parity unambiguously \cite{Wang1508-EPJC,WangHuang-EPJC-1508-12,WangZG-EPJC-1509-12,WangZG-NPB-1512-32,
WangZhang-APPB}.

The quark-hadron duality allows the correlation functions at the quark level to be equivalent to those at the hadron level within a certain energy scale \cite{Weinberg-1976}. The currents $J(x)$, $J_{\mu}(x)$ and $J_{\mu\nu}(x)$ might also couple potentially  to the meson-baryon scattering states $\bar{D}_s\Xi_c$, $\bar{D}^*_s\Xi_c$, $\bar{D}\Omega_c$, $\bar{D}^*\Omega_c$,
$J/\psi \Xi$, $\eta_c \Xi$, $\cdots$, if they have the same quantum numbers.
Thus there are intermediate meson-baryon loops (in other words, self-energies) beyond the pole terms,
direct calculations indicate that the meson-baryon scattering states alone cannot saturate the QCD sum rules,
 the renormalized self-energies  contribute  a finite imaginary part $\sqrt{p^2}\Gamma(p^2)$ to modify the dispersion relation \cite{WangZG-Review,WangZG-landau-PRD},
\begin{eqnarray}\label{Modify-width}
\Pi_{j}^{1/0}(p) &=&-\frac{\lambda_{P}^{2}/\lambda_{P}^{2}M_P}{ p^2-M_{P}^2+i\sqrt{p^2}\Gamma(p^2)}+\cdots \, ,
 \end{eqnarray}
 with $j=\frac{1}{2}$, $\frac{3}{2}$, $\frac{5}{2}$.
We  take  account of the finite width via a simple replacement,
\begin{eqnarray}
\delta \left(s-M^2_{P} \right) &\to& \frac{1}{\pi}\frac{M_{P}\Gamma_{P}}{(s-M_{P}^2)^2+M_{P}^2\Gamma_{P}^2}\, ,
\end{eqnarray}
with  $\Gamma(M_P^2)=\Gamma_P$.
Then the hadron  sides of  the QCD sum rules undergo the change,
\begin{eqnarray}
\lambda^2_{P}\exp \left(-\frac{M^2_{P}}{T^2} \right) &\to& \lambda^2_{P}\int_{(m_{\Xi}+m_{\eta_c})^2}^{s_0}
ds\frac{1}{\pi}\frac{M_{P}\Gamma_{P}}{(s-M_{P}^2)^2+M_{P}^2\Gamma_{P}^2}\exp \left(-\frac{s}{T^2} \right)\, , \nonumber\\
&=&C_P^2\,\lambda^2_{P}\exp \left(-\frac{M^2_{P}}{T^2} \right)\, .
\end{eqnarray}
We  can absorb the numerical factors  $C_P$ into the pole residues $\lambda_P$ safely, the intermediate   meson-baryon loops cannot  affect  the masses  $M_{P}$ significantly \cite{WangZG-Review,WangZG-landau-PRD,WangZG-IJMPA-Two-part}.
All in all, we should bear in mind that we choose the local currents, and the traditional mesons and baryons are spatial extended objects and have mean spatial sizes $\sqrt{\langle r^2\rangle} \neq 0$,  the overlaps of the wave-functions are very small indeed \cite{WangZG-Review}.

On the QCD side,  we accomplish  the operator product expansion resort to the full $u$, $d$, $s$ and $c$ quark propagators,
 \begin{eqnarray}
U/D_{ij}(x)&=& \frac{i\delta_{ij}\!\not\!{x}}{ 2\pi^2x^4}-\frac{\delta_{ij}\langle
\bar{q}q\rangle}{12} -\frac{\delta_{ij}x^2\langle \bar{q}g_s\sigma Gq\rangle}{192} -\frac{ig_sG^{a}_{\alpha\beta}t^a_{ij}(\!\not\!{x}
\sigma^{\alpha\beta}+\sigma^{\alpha\beta} \!\not\!{x})}{32\pi^2x^2} -\frac{\delta_{ij}x^4\langle \bar{q}q \rangle\langle g_s^2 GG\rangle}{27648} \nonumber\\
&&  -\frac{1}{8}\langle\bar{q}_j\sigma^{\mu\nu}q_i \rangle \sigma_{\mu\nu}+\cdots \, ,
\end{eqnarray}
\begin{eqnarray}
S_{ij}(x)&=& \frac{i\delta_{ij}\!\not\!{x}}{ 2\pi^2x^4}
-\frac{\delta_{ij}m_s}{4\pi^2x^2}-\frac{\delta_{ij}\langle
\bar{s}s\rangle}{12} +\frac{i\delta_{ij}\!\not\!{x}m_s
\langle\bar{s}s\rangle}{48}-\frac{\delta_{ij}x^2\langle \bar{s}g_s\sigma Gs\rangle}{192}+\frac{i\delta_{ij}x^2\!\not\!{x} m_s\langle \bar{s}g_s\sigma
 Gs\rangle }{1152}\nonumber\\
&& -\frac{ig_s G^{a}_{\alpha\beta}t^a_{ij}(\!\not\!{x}
\sigma^{\alpha\beta}+\sigma^{\alpha\beta} \!\not\!{x})}{32\pi^2x^2} -\frac{\delta_{ij}x^4\langle \bar{s}s \rangle\langle g_s^2 GG\rangle}{27648}-\frac{1}{8}\langle\bar{s}_j\sigma^{\mu\nu}s_i \rangle \sigma_{\mu\nu}  +\cdots \, ,
\end{eqnarray}
\begin{eqnarray}
C_{ij}(x)&=&\frac{i}{(2\pi)^4}\int d^4k e^{-ik \cdot x} \left\{
\frac{\delta_{ij}}{\!\not\!{k}-m_c}
-\frac{g_sG^n_{\alpha\beta}t^n_{ij}}{4}\frac{\sigma^{\alpha\beta}(\!\not\!{k}+m_c)+(\!\not\!{k}+m_c)
\sigma^{\alpha\beta}}{(k^2-m_c^2)^2}\right.\nonumber\\
&&\left. -\frac{g_s^2 (t^at^b)_{ij} G^a_{\alpha\beta}G^b_{\mu\nu}(f^{\alpha\beta\mu\nu}+f^{\alpha\mu\beta\nu}+f^{\alpha\mu\nu\beta}) }{4(k^2-m_c^2)^5}+\cdots\right\} \, ,\nonumber\\
f^{\alpha\beta\mu\nu}&=&(\!\not\!{k}+m_c)\gamma^\alpha(\!\not\!{k}+m_c)\gamma^\beta(\!\not\!{k}+m_c)\gamma^\mu(\!\not\!{k}+m_c)\gamma^\nu(\!\not\!{k}+m_c)\, ,
\end{eqnarray}
and  $t^n=\frac{\lambda^n}{2}$, the $\lambda^n$ is the Gell-Mann matrix
\cite{PRT85,Pascual-1984,WangHuang3900}.
We introduce  the operators $\langle\bar{q}_j\sigma_{\mu\nu}q_i \rangle$ and $\langle\bar{s}_j\sigma_{\mu\nu}s_i \rangle$ originate  from Fierz transformations   of the quark-antiquark pairs
$\langle q_i \bar{q}_j\rangle$ and $\langle s_i \bar{s}_j\rangle$  to  absorb the gluons  emitted from other quark lines to  extract the mixed condensates   $\langle\bar{q}g_s\sigma G q\rangle$ and $\langle\bar{s}g_s\sigma G s\rangle$, respectively \cite{WangHuang3900}.  Then we calculate all the Feynman diagrams to obtain  the QCD spectral densities through   dispersion relation,
\begin{eqnarray}\label{QCD-rho}
 \rho^1_{QCD}(s) &=&\frac{{\rm Im}\Pi^1_j(s)}{\pi}\, , \nonumber\\
\rho^0_{QCD}(s) &=&\frac{{\rm Im}\Pi^0_j(s)}{\pi}\, ,
\end{eqnarray}
where $j=\frac{1}{2}$, $\frac{3}{2}$, $\frac{5}{2}$.
Just as in our previous works \cite{WangZG-Review,WZG-penta-IJMPA,WangZG-Pcs4459-333,Wang-tetra-PRD-HC,
Wang-tetra-NPB-HCss,WangZG-IJMPA-2021,WZG-tetraquark-Mc}, we take account of the quark-gluon operators up to dimension $13$ and order $\mathcal{O}( \alpha_s^{k})$ with $k\leq 1$ in a consistent way to select the vacuum condensates, and   take account of the terms  $\propto m_s$ to embody  the light-flavor   $SU(3)$ mass-breaking effects, which  is our unique counting scheme \cite{WangZG-Review}.

Now we  match the hadron side with the QCD side of the components $\Pi_{\frac{1}{2}}^{1/0}(p^2)$, $\Pi_{\frac{3}{2}}^{1/0}(p^2)$ and  $\Pi_{\frac{5}{2}}^{1/0}(p^2)$, take the quark-hadron duality below the continuum thresholds, and  obtain  two  QCD sum rules:
\begin{eqnarray}\label{QCDSR}
2M_{-}\lambda_{-}^2\exp\left( -\frac{M_{-}^2}{T^2}\right)&=& \int_{4m_c^2}^{s_0}ds \,\left[\sqrt{s}\rho_{QCD}^1(s)+\rho_{QCD}^{0}(s)\right]\,\exp\left( -\frac{s}{T^2}\right)\,  ,
\end{eqnarray}
\begin{eqnarray}\label{QCDSR-Positive}
2M_{+}\lambda_{+}^2\exp\left( -\frac{M_{+}^2}{T^2}\right)&=& \int_{4m_c^2}^{s^\prime_0}ds \,\left[\sqrt{s}\rho_{QCD}^1(s)-\rho_{QCD}^{0}(s)\right]\,\exp\left( -\frac{s}{T^2}\right)\,  .
\end{eqnarray}

We usually obtain two traditional QCD sum rules with respect to the components $\Pi_{\frac{1}{2}/\frac{3}{2}/\frac{5}{2}}^{1}(p^2)$ and $\Pi_{\frac{1}{2}/\frac{3}{2}/\frac{5}{2}}^{0}(p^2)$, respectively \cite{PRT85},
\begin{eqnarray}\label{Traditional-QCDSR-1}
\lambda_{-}^2\exp\left( -\frac{M_{-}^2}{T^2}\right)+\lambda_{+}^2\exp\left( -\frac{M_{+}^2}{T^2}\right)&=&\int_{4m_c^2}^{s_0}ds \,\rho^1_{QCD}(s)\exp\left( -\frac{s}{T^2}\right) \, ,
\end{eqnarray}
\begin{eqnarray}\label{Traditional-QCDSR-0}
M_{-}\lambda_{-}^2\exp\left( -\frac{M_{-}^2}{T^2}\right)-M_{+}\lambda_{+}^2\exp\left( -\frac{M_{+}^2}{T^2}\right)&=&\int_{4m_c^2}^{s_0}ds \,\rho^0_{QCD}(s)\exp\left( -\frac{s}{T^2}\right) \, ,
\end{eqnarray}
if we also take account of the contributions of the $P_{css}$ states with  positive parity. In the case that we set the hadronic coupling constants  $\lambda_{+}$ to be zero, the contaminations from the $P_{css}$ states with  positive parity are included in the masses $M_{-}$ and pole residues $\lambda_{-}$.
 In Refs.\cite{WangZG-Pc12-Jpsip,WangZG-Pc12-JpsiLambda},  we define a parameter CTM to measure the contaminations from the hidden-charm-doubly-strange  pentaquark states $P^{+}_{css}$,
\begin{eqnarray}
{\rm CTM}&=&\frac{\int_{4m_c^2}^{s_0}ds \,\left[\sqrt{s}\rho_{QCD}^1(s)-\rho_{QCD}^{0}(s)\right]\,\exp\left( -\frac{s}{T^2}\right)}{\int_{4m_c^2}^{s_0}ds \,\left[\sqrt{s}\rho_{QCD}^1(s)+\rho_{QCD}^{0}(s)\right]\,\exp\left( -\frac{s}{T^2}\right)}\, , \nonumber\\
&=&\frac{M_{+}\lambda_{+}^2\exp\left( -\frac{M_{+}^2}{T^2}\right)}{M_{-}\lambda_{-}^2\exp\left( -\frac{M_{-}^2}{T^2}\right)}\, ,
\end{eqnarray}
by setting $s^\prime_0=s_0$.
Direct calculations indicate that ${\rm CTM}\sim 0.10$ or $0.20$ in the Borel windows, the contaminations from the hidden-charm-doubly-strange pentaquark states with positive parity are rather large. We should bear in mind that the values of the $M_{+}$ and $\lambda_{+}$ from the QCD sum rules with the continuum threshold parameters $s^\prime_0=s_0$ are smaller than the true values, as the physical masses $M_{+}> M_{-}$ due to the additional P-wave in the $P_{css}$ state with positive parity.

In the present studies, we adopt the QCD sum rules for the hidden-charm-doubly-strange pentaquark states $P_{css}^{-}$ with negative parity, and differentiate   Eq.\eqref{QCDSR} with respect   to  $\frac{1}{T^2}$, then eliminate the pole residues $\lambda_{-}$ to get  the QCD sum rules for the  masses,
 \begin{eqnarray}
 M^2_{-} &=& \frac{-\int_{4m_c^2}^{s_0}ds \frac{d}{d(1/T^2)}\, \left[\sqrt{s}\rho_{QCD}^1(s)+\rho_{QCD}^{0}(s)\right]\,\exp\left( -\frac{s}{T^2}\right)}{\int_{4m_c^2}^{s_0}ds \, \left[\sqrt{s}\rho_{QCD}^1(s)+\rho_{QCD}^{0}(s)\right]\,\exp\left( -\frac{s}{T^2}\right)}\,  .
\end{eqnarray}
Thereafter, we would like to neglect the subscript $-$ for simplicity.

\section{Numerical results and discussions}
At the beginning points, we take  the standard values of the  vacuum condensates
$\langle\bar{q}q \rangle=-(0.24\pm 0.01\, \rm{GeV})^3$,  $\langle\bar{s}s \rangle=(0.8\pm0.1)\langle\bar{q}q \rangle$,
 $\langle\bar{q}g_s\sigma G q \rangle=m_0^2\langle \bar{q}q \rangle$, $\langle\bar{s}g_s\sigma G s \rangle=m_0^2\langle \bar{s}s \rangle$,
$m_0^2=(0.8 \pm 0.1)\,\rm{GeV}^2$, $\langle \frac{\alpha_s
GG}{\pi}\rangle=0.012\pm0.004\,\rm{GeV}^4$    at the energy scale  $\mu=1\, \rm{GeV}$
\cite{SVZ79-1,SVZ79-2,PRT85,ColangeloReview}, and  take the $\overline{MS}$ (modified-minimal-subtraction) quark  masses $m_{c}(m_c)=(1.275\pm0.025)\,\rm{GeV}$
 and $m_s(\mu=2\,\rm{GeV})=(0.095\pm0.005)\,\rm{GeV}$
 from the Particle Data Group \cite{PDG}.
Moreover,  we consider  the energy-scale dependence of  those  input parameters  \cite{Narison-mix},
 \begin{eqnarray}
 \langle\bar{q}q \rangle(\mu)&=&\langle\bar{q}q\rangle({\rm 1 GeV})\left[\frac{\alpha_{s}({\rm 1 GeV})}{\alpha_{s}(\mu)}\right]^{\frac{12}{33-2n_f}}\, , \nonumber\\
 \langle\bar{s}s \rangle(\mu)&=&\langle\bar{s}s \rangle({\rm 1 GeV})\left[\frac{\alpha_{s}({\rm 1 GeV})}{\alpha_{s}(\mu)}\right]^{\frac{12}{33-2n_f}}\, , \nonumber\\
 \langle\bar{q}g_s \sigma Gq \rangle(\mu)&=&\langle\bar{q}g_s \sigma Gq \rangle({\rm 1 GeV})\left[\frac{\alpha_{s}({\rm 1 GeV})}{\alpha_{s}(\mu)}\right]^{\frac{2}{33-2n_f}}\, ,\nonumber\\
  \langle\bar{s}g_s \sigma Gs \rangle(\mu)&=&\langle\bar{s}g_s \sigma Gs \rangle({\rm 1 GeV})\left[\frac{\alpha_{s}({\rm 1 GeV})}{\alpha_{s}(\mu)}\right]^{\frac{2}{33-2n_f}}\, ,\nonumber\\
m_c(\mu)&=&m_c(m_c)\left[\frac{\alpha_{s}(\mu)}{\alpha_{s}(m_c)}\right]^{\frac{12}{33-2n_f}} \, ,\nonumber\\
m_s(\mu)&=&m_s({\rm 2GeV} )\left[\frac{\alpha_{s}(\mu)}{\alpha_{s}({\rm 2GeV})}\right]^{\frac{12}{33-2n_f}}\, ,\nonumber\\
\alpha_s(\mu)&=&\frac{1}{b_0t}\left[1-\frac{b_1}{b_0^2}\frac{\log t}{t} +\frac{b_1^2(\log^2{t}-\log{t}-1)+b_0b_2}{b_0^4t^2}\right]\, ,
\end{eqnarray}
  where $t=\log \frac{\mu^2}{\Lambda^2}$, $b_0=\frac{33-2n_f}{12\pi}$, $b_1=\frac{153-19n_f}{24\pi^2}$, $b_2=\frac{2857-\frac{5033}{9}n_f+\frac{325}{27}n_f^2}{128\pi^3}$,  $\Lambda_{QCD}=210\,\rm{MeV}$, $292\,\rm{MeV}$  and  $332\,\rm{MeV}$ for the flavors  $n_f=5$, $4$ and $3$, respectively  \cite{PDG}.

In the present studies, we explore  the hidden-charm-doubly-strange  pentaquark states $qssc\bar{c}$ with the isospin $I=\frac{1}{2}$,  and choose the total flavor numbers $n_f=4$, then evolve all those input parameters to a typical energy scale $\mu$, which satisfies  the modified energy scale formula,
\begin{eqnarray}
\mu &=&\sqrt{M_{P}^2-(2{\mathbb{M}}_c)^2}-2{\mathbb{M}}_s \, ,
 \end{eqnarray}
 with the effective quark masses ${\mathbb{M}}_c$ and ${\mathbb{M}}_s$, which embody the heavy degrees of freedom and light-flavor $SU(3)$ mass-breaking effects, respectively,  the  updated values are ${\mathbb{M}}_c=1.82\,\rm{GeV}$ and ${\mathbb{M}}_s=0.15\,\rm{GeV}$ respectively, and the energy scale formula works well for the hidden-charm/bottom (doubly-charm/bottom) tetraquark states, pentaquark states and molecular states \cite{Pcs4338-mole-XWWang,WangZG-Review,WangZG-Pcs4459-333,WangZG-Pc12-JpsiLambda,
 Wang-tetra-PRD-HC,Wang-tetra-NPB-HCss,WangZG-IJMPA-2021,
 WZG-tetraquark-Mc,Wang-tetra-formula,WangZG-mole-formula-1,WangZG-mole-formula-2,
WangZG-XQ-mole-EPJA}.

In the QCD sum rules for  the  baryons  and  pentaquark (molecular) states contain at least one valence heavy quark,  we usually choose the continuum threshold parameters as $\sqrt{s_0}=M_{gr}+ (0.5-0.8)\,\rm{GeV}$  \cite{Pc4312-mole-penta-WXW-SCPMA,
Pc4312-mole-penta-WXW-IJMPA,Pcs4338-mole-XWWang,WangZG-Review,Wang1508-EPJC,WangHuang-EPJC-1508-12,
 WangZG-EPJC-1509-12,WangZG-NPB-1512-32,WZG-penta-IJMPA,WangZG-Pcs4459-333,
 WangZG-Pc12-Jpsip,WangZG-Pc12-JpsiLambda,
 Wang-cc-baryon-penta},   where the subscript $gr$ stands for the ground states.
 In our previous works \cite{WangZG-Pc12-Jpsip,WangZG-Pc12-JpsiLambda},  we adopted such a criterion and studied the mass spectrum of the $udsc\bar{c}$ and $uudc\bar{c}$ pentaquark states with the isospins $I=0$ and $\frac{1}{2}$ respectively in a comprehensive way,  and made possible assignments of the $P_c(4312)$, $P_c(4337)$, $P_{cs}(4338)$, $P_{c}(4380)$, $P_c(4440)$, $P_c(4457)$ and $P_{cs}(4459)$ in a suitable way. Now we extend our previous works to study the lowest  hidden-charm-double-strange pentaquark states $P_{css}$ with the  isospin $I=\frac{1}{2}$ in the $J/\psi \Xi$ mass spectrum.

 We obtain the  Borel  windows and continuum threshold parameters via numerous  trial  and error according to the  four   criteria:\\
$\bullet$ Pole dominance on the hadron  side;\\
$\bullet$ Convergence of the operator product expansion on the QCD side;\\
$\bullet$ Appearance of the Borel platforms;\\
$\bullet$ Satisfying the modified  energy scale formula,\\
 and  the numerical results are  shown explicitly in Table \ref{Borel}. From the table, we can see explicitly  that the pole contributions are about $(40-60)\%$, the pole dominance criterion is satisfied, where
  the pole contributions are defined by,
\begin{eqnarray}
{\rm{pole}}&=&\frac{\int_{4m_{c}^{2}}^{s_{0}}ds\,\rho_{QCD}\left(s\right)\exp\left(-\frac{s}{T^{2}}\right)} {\int_{4m_{c}^{2}}^{\infty}ds\,\rho_{QCD}\left(s\right)\exp\left(-\frac{s}{T^{2}}\right)}\, ,
\end{eqnarray}
 with $\rho_{QCD}=\sqrt{s}\rho_{QCD}^1(s)+\rho_{QCD}^{0}(s)$.

\begin{figure}
\centering
\includegraphics[totalheight=6cm,width=7cm]{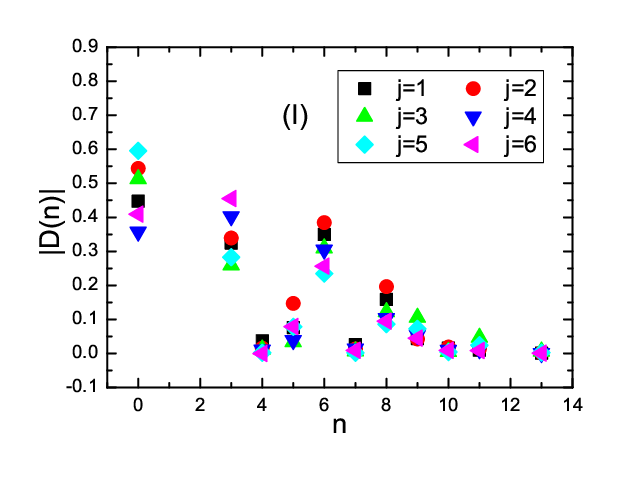}
\includegraphics[totalheight=6cm,width=7cm]{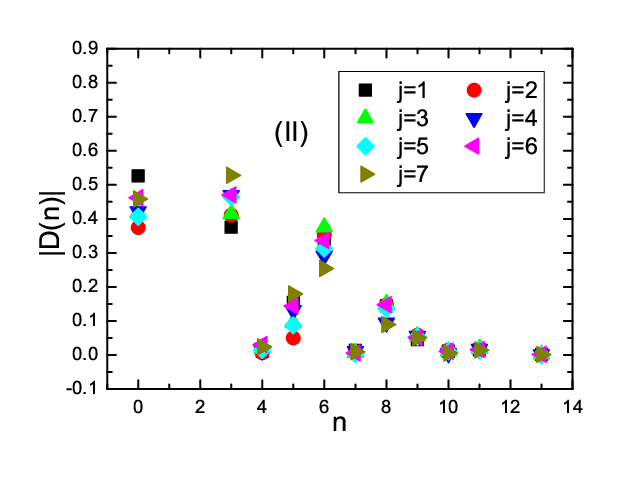}
\includegraphics[totalheight=6cm,width=7cm]{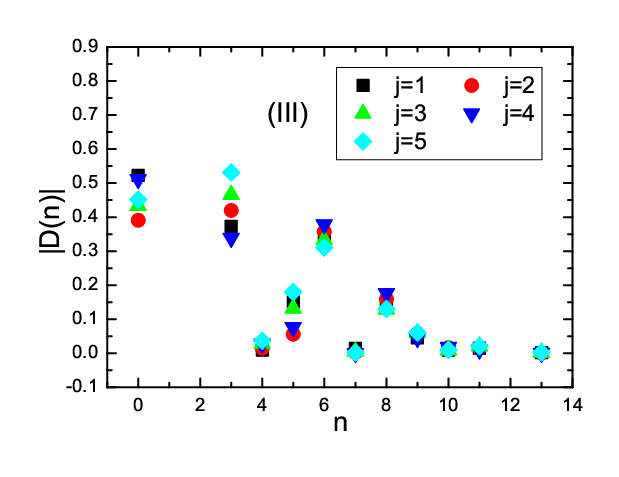}
 \caption{ The $|D(n)|$ with variations of the $n$ for the central values of the input parameters, where the (I), (II) and (III) denote  the spins  $J=\frac{1}{2}$, $\frac{3}{2}$ and $\frac{5}{2}$ of the currents respectively, the $j=1$, $2$, $3$, $4$, $5$, $6$ and $7$ denote the series numbers of the currents. }\label{OPE-fig}
\end{figure}

 In Fig.\ref{OPE-fig}, we plot the absolute values of the contributions of the vacuum condensates with the dimension $n$ for the central values of all the other  parameters, where the $D(n)$ are defined by,
   \begin{eqnarray}
D(n)&=&\frac{\int_{4m_{c}^{2}}^{s_{0}}ds\,\rho_{QCD,n}(s)\exp\left(-\frac{s}{T^{2}}\right)}
{\int_{4m_{c}^{2}}^{s_{0}}ds\,\rho_{QCD}\left(s\right)\exp\left(-\frac{s}{T^{2}}\right)}\, .
\end{eqnarray}
From the figure, we can see explicitly  that the $D(4)$ and $D(7)$ play a tiny role, while the $D(6)$ plays  a most  important role, and serves as a milestone. The contributions $|D(n)|$ of the vacuum condensates with $n\geq 6$ have the hierarchies,
\begin{eqnarray}
&&D(6)\gg |D(8)| \gg D(9) \gg D(10)\sim |D(11)| \gg D(13) \, ,
\end{eqnarray}
in general,  the operator product expansion converges  very good.

\begin{table}
\begin{center}
\begin{tabular}{|c|c|c|c|c|c|c|c|}\hline\hline
                  &$T^2(\rm{GeV}^2)$     &$\sqrt{s_0}(\rm{GeV})$    &$\mu(\rm{GeV})$  &pole          &$D(13)$         \\ \hline

$J^1(x)$          &$3.6-4.0$             &$5.30\pm0.10$             &$2.5$            &$(41-60)\%$   &$\ll 1\%$      \\ \hline

$J^2(x)$          &$3.6-4.0$             &$5.40\pm0.10$             &$2.7$            &$(42-62)\%$   &$\ll 1\%$       \\ \hline

$J^3(x)$          &$3.2-3.6$             &$5.18\pm0.10$             &$2.3$            &$(41-63)\%$   &$< 1\%$      \\ \hline

$J^4(x)$          &$3.4-3.8$             &$5.17\pm0.10$             &$2.3$            &$(41-61)\%$   &$\ll1\%$     \\ \hline

$J^5(x)$          &$3.5-3.9$             &$5.35\pm0.10$             &$2.6$            &$(40-60)\%$   &$\ll 1\%$      \\ \hline

$J^6(x)$          &$3.6-4.0$             &$5.30\pm0.10$             &$2.5$            &$(40-60)\%$   &$\ll1\%$     \\ \hline

$J^1_\mu(x)$      &$3.7-4.1$             &$5.40\pm0.10$             &$2.7$            &$(41-60)\%$   &$\ll 1\%$     \\ \hline

$J^2_\mu(x)$      &$3.5-3.9$             &$5.20\pm0.10$             &$2.4$            &$(40-60)\%$   &$\ll1\%$     \\ \hline

$J^3_\mu(x)$      &$3.5-3.9$             &$5.25\pm0.10$             &$2.4$            &$(41-61)\%$   &$\ll1\%$     \\ \hline

$J^4_\mu(x)$      &$3.5-3.9$             &$5.25\pm0.10$             &$2.4$            &$(40-60)\%$   &$\ll1\%$     \\ \hline

$J^5_\mu(x)$      &$3.6-4.0$             &$5.30\pm0.10$             &$2.5$            &$(41-61)\%$   &$\ll1\%$     \\ \hline

$J^6_\mu(x)$      &$3.6-4.0$             &$5.35\pm0.10$             &$2.6$            &$(42-61)\%$   &$\ll1\%$     \\ \hline

$J^7_\mu(x)$      &$3.6-4.0$             &$5.35\pm0.10$             &$2.6$            &$(41-61)\%$   &$\ll1\%$     \\ \hline

$J^1_{\mu\nu}(x)$ &$3.7-4.1$             &$5.40\pm0.10$             &$2.7$            &$(42-61)\%$   &$\ll1\%$     \\ \hline

$J^2_{\mu\nu}(x)$ &$3.5-3.9$             &$5.23\pm0.10$             &$2.4$            &$(41-61)\%$   &$\ll1\%$     \\ \hline

$J^3_{\mu\nu}(x)$ &$3.6-4.0$             &$5.29\pm0.10$             &$2.5$            &$(41-60)\%$   &$\ll1\%$     \\ \hline

$J^4_{\mu\nu}(x)$ &$3.7-4.1$             &$5.40\pm0.10$             &$2.7$            &$(42-61)\%$   &$\ll1\%$     \\ \hline

$J^5_{\mu\nu}(x)$ &$3.6-4.0$             &$5.35\pm0.10$             &$2.6$            &$(42-62)\%$   &$\ll1\%$     \\ \hline

\hline
\end{tabular}
\end{center}
\caption{ The Borel  windows, continuum threshold parameters, ideal energy scales, pole contributions,   contributions of the vacuum condensates of dimension 13 for the hidden-charm-doubly-strange pentaquark states with the isospin $I=\frac{1}{2}$. }\label{Borel}
\end{table}

\begin{table}
\begin{center}
\begin{tabular}{|c|c|c|c|c|c|c|c|c|}\hline\hline
$[qq][qc]\bar{c}$ ($S_L$, $S_H$, $J_{LH}$, $J$) &$M(\rm{GeV})$   &$\lambda(10^{-3}\rm{GeV}^6)$       \\ \hline

$[qs][sc]\bar{c}$ ($0$, $0$, $0$, $\frac{1}{2}$)  &$4.61\pm0.11$ &$2.32\pm0.36$                  \\ \hline

$[qs][sc]\bar{c}$ ($0$, $1$, $1$, $\frac{1}{2}$)  &$4.71\pm0.10$ &$4.60\pm0.71$                 \\ \hline

$[ss][qc]\bar{c}-[sq][sc]\bar{c}$ ($1$, $1$, $0$, $\frac{1}{2}$)  &$4.49\pm0.12$ &$3.77\pm0.66$                \\ \hline

$[ss][qc]\bar{c}-[sq][sc]\bar{c}$ ($1$, $0$, $0$, $\frac{1}{2}$)  &$4.48\pm0.11$ &$4.00\pm0.66$                \\ \hline

$[ss][qc]\bar{c}$ ($1$, $1$, $0$, $\frac{1}{2}$)  &$4.65\pm0.11$ &$5.94\pm1.00$                \\ \hline

$[ss][qc]\bar{c}$ ($1$, $0$, $0$, $\frac{1}{2}$)  &$4.61\pm0.11$ &$5.71\pm0.90$                \\ \hline

$[qs][sc]\bar{c}$ ($0$, $1$, $1$, $\frac{3}{2}$)  &$4.70\pm0.10$ &$2.54\pm0.39$                \\ \hline

$[ss][qc]\bar{c}-[sq][sc]\bar{c}$ ($1$, $0$, $1$, $\frac{3}{2}$)  &$4.51\pm0.10$ &$2.33\pm0.37$                \\ \hline

$[ss][qc]\bar{c}-[sq][sc]\bar{c}$ ($1$, $1$, $2$, $\frac{3}{2}$)${}_3$ &$4.56\pm0.10$  &$4.24\pm0.67$   \\ \hline

$[ss][qc]\bar{c}-[sq][sc]\bar{c}$ ($1$, $1$, $2$, $\frac{3}{2}$)${}_4$ &$4.56\pm0.11$   &$4.25\pm0.69$    \\ \hline

$[ss][qc]\bar{c}$ ($1$, $0$, $1$, $\frac{3}{2}$)  &$4.61\pm0.11$ &$3.10\pm0.48$                \\ \hline

$[ss][qc]\bar{c}$ ($1$, $1$, $2$, $\frac{3}{2}$)${}_6$ &$4.66\pm0.11$  &$5.74\pm0.89$   \\ \hline

$[ss][qc]\bar{c}$ ($1$, $1$, $2$, $\frac{3}{2}$)${}_7$ &$4.66\pm0.11$   &$5.75\pm0.90$    \\ \hline

$[qs][sc]\bar{c}$ ($0$, $1$, $1$, $\frac{5}{2}$)  &$4.70\pm0.10$ &$2.54\pm0.38$                     \\ \hline

$[ss][qc]\bar{c}-[sq][sc]\bar{c}$ ($1$, $0$, $1$, $\frac{5}{2}$)  &$4.54\pm0.10$ &$2.40\pm0.38$                     \\ \hline

$[ss][qc]\bar{c}-[sq][sc]\bar{c}$ ($1$, $1$, $2$, $\frac{5}{2}$)  &$4.60\pm0.11$   &$2.49\pm0.39$                  \\ \hline

$[qs][sc]\bar{c}$ ($1$, $0$, $1$, $\frac{5}{2}$)  &$4.69\pm0.10$ &$2.54\pm0.38$                     \\ \hline

$[ss][qc]\bar{c}$ ($1$, $1$, $2$, $\frac{5}{2}$)  &$4.66\pm0.11$   &$3.12\pm0.48$                  \\ \hline\hline
\end{tabular}
\end{center}
\caption{ The masses  and pole residues of the hidden-charm-doubly-strange pentaquark states. }\label{mass-Pcs}
\end{table}

\begin{figure}
\centering
\includegraphics[totalheight=6cm,width=7cm]{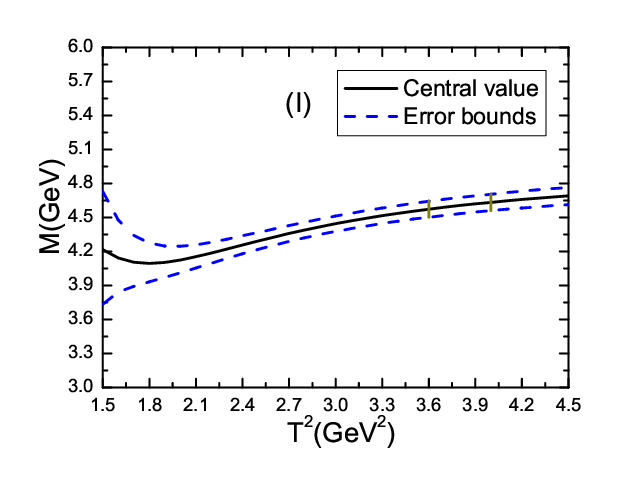}
\includegraphics[totalheight=6cm,width=7cm]{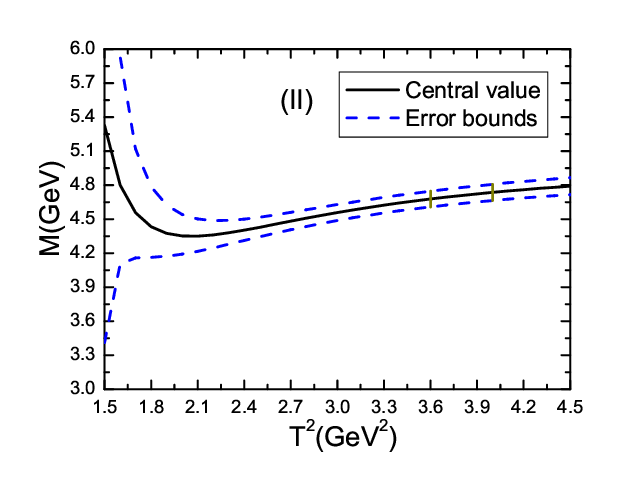}
\includegraphics[totalheight=6cm,width=7cm]{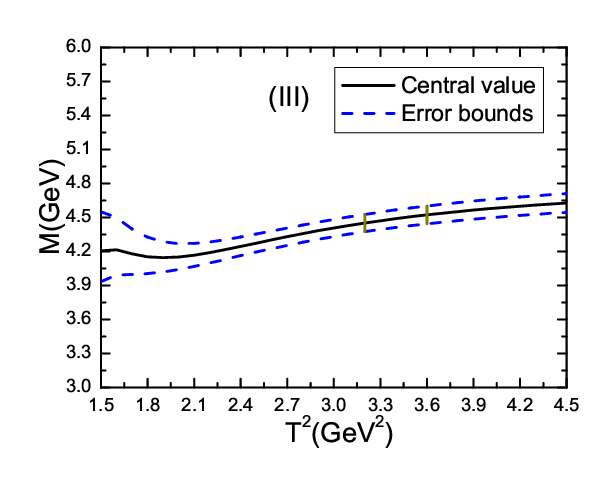}
\includegraphics[totalheight=6cm,width=7cm]{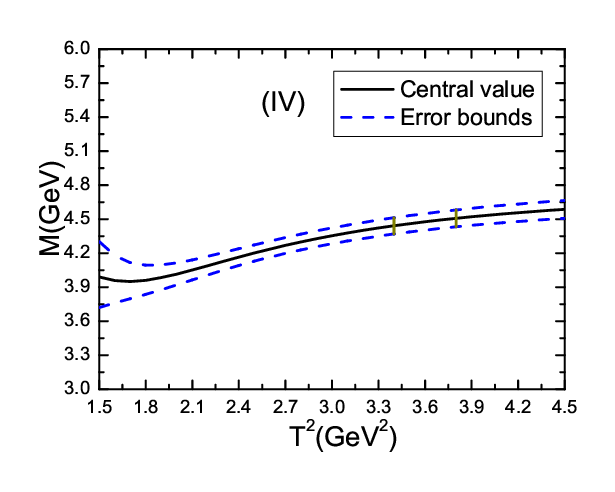}
\includegraphics[totalheight=6cm,width=7cm]{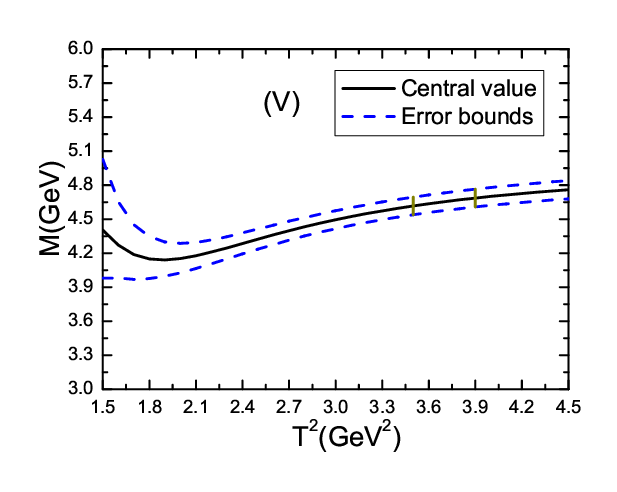}
\includegraphics[totalheight=6cm,width=7cm]{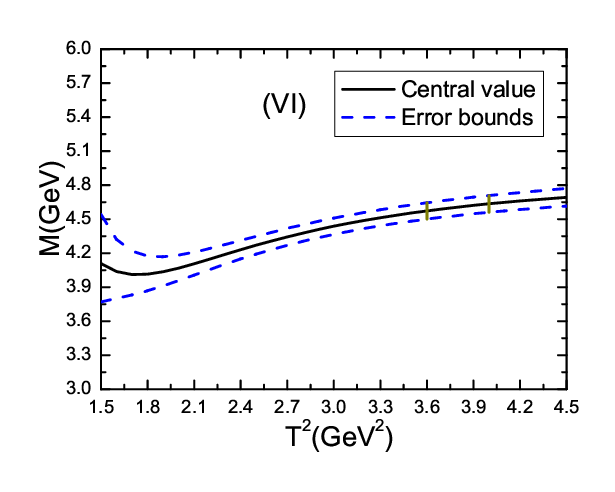}
  \caption{ The masses  with variations of the  Borel parameters $T^2$ for  the hidden-charm-doubly-strange  pentaquark states, where the (I), (II), (III), (IV), (V)  and (VI)  denote the
   $[qs][sc]\bar{c}$ ($0$, $0$, $0$, $\frac{1}{2}$),                    $[qs][sc]\bar{c}$ ($0$, $1$, $1$, $\frac{1}{2}$), $[ss][qc]\bar{c}-[sq][sc]\bar{c}$ ($1$, $1$, $0$, $\frac{1}{2}$),
$[ss][qc]\bar{c}-[sq][sc]\bar{c}$ ($1$, $0$, $0$, $\frac{1}{2}$),
$[ss][qc]\bar{c}$ ($1$, $1$, $0$, $\frac{1}{2}$)   and
$[ss][qc]\bar{c}$ ($1$, $0$, $0$, $\frac{1}{2}$)  pentaquark states, respectively. }\label{mass-1-fig}
\end{figure}

\begin{figure}
\centering
\includegraphics[totalheight=6cm,width=7cm]{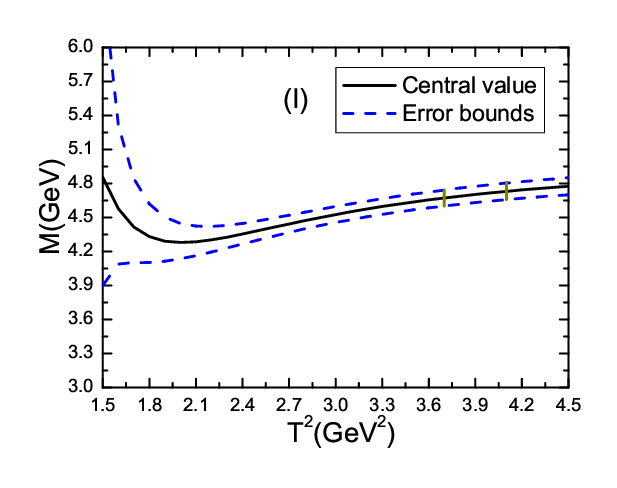}
\includegraphics[totalheight=6cm,width=7cm]{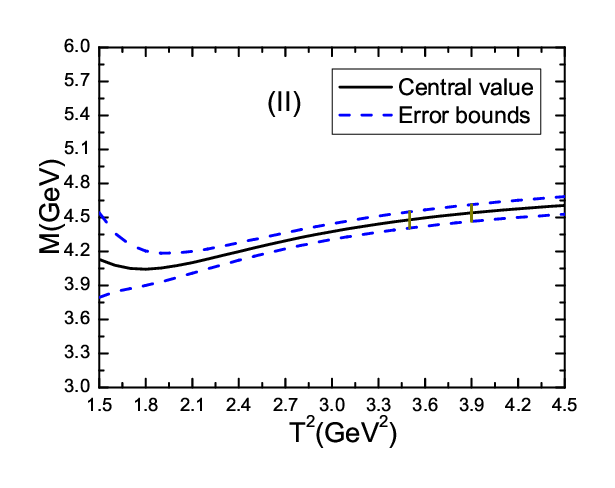}
\includegraphics[totalheight=6cm,width=7cm]{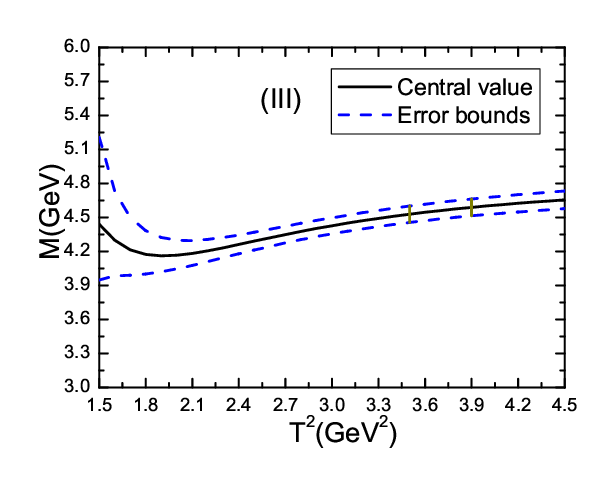}
\includegraphics[totalheight=6cm,width=7cm]{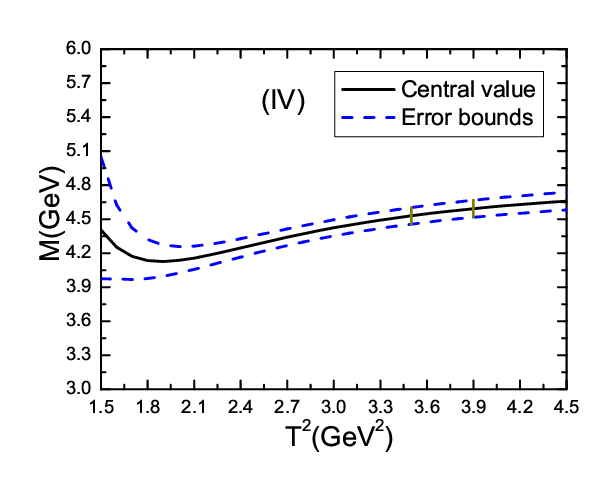}
\includegraphics[totalheight=6cm,width=7cm]{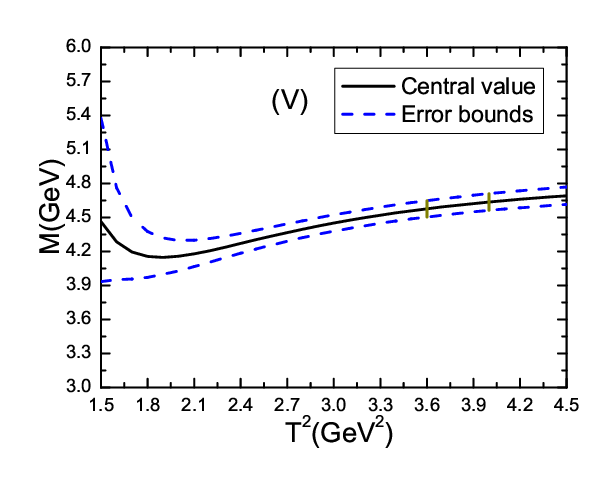}
\includegraphics[totalheight=6cm,width=7cm]{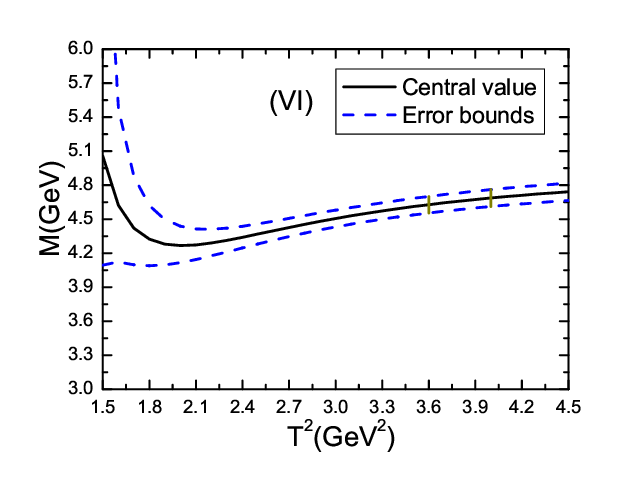}
  \caption{ The masses  with variations of the  Borel parameters $T^2$ for  the hidden-charm-doubly-strange pentaquark states, where the (I), (II), (III), (IV), (V) and (VI) denote the $[qs][sc]\bar{c}$ ($0$, $1$, $1$, $\frac{3}{2}$),
$[ss][qc]\bar{c}-[sq][sc]\bar{c}$ ($1$, $0$, $1$, $\frac{3}{2}$),
$[ss][qc]\bar{c}-[sq][sc]\bar{c}$ ($1$, $1$, $2$, $\frac{3}{2}$)${}_3$,
$[ss][qc]\bar{c}-[sq][sc]\bar{c}$ ($1$, $1$, $2$, $\frac{3}{2}$)${}_4$, $[ss][qc]\bar{c}$ ($1$, $0$, $1$, $\frac{3}{2}$) and $[ss][qc]\bar{c}$ ($1$, $1$, $2$, $\frac{3}{2}$)${}_6$  pentaquark states, respectively.  }\label{mass-2-fig-1}
\end{figure}

\begin{figure}
\centering
\includegraphics[totalheight=6cm,width=7cm]{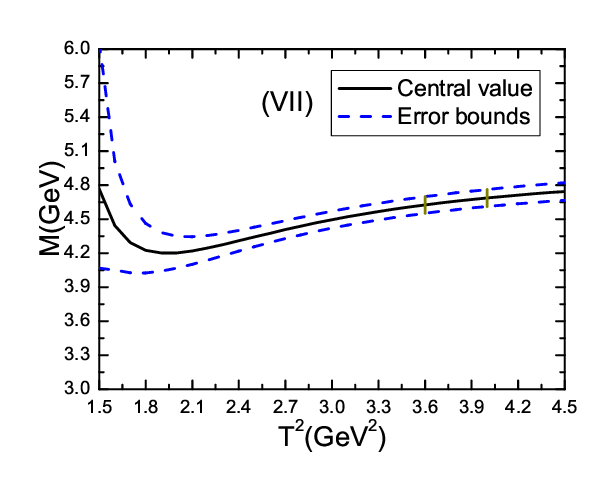}
  \caption{ The mass  with variations of the  Borel parameter $T^2$ for  the hidden-charm-doubly-strange pentaquark state, where the (VII) denotes the
$[ss][qc]\bar{c}$ ($1$, $1$, $2$, $\frac{3}{2}$)${}_7$ pentaquark state.  }\label{mass-2-fig-2}
\end{figure}

\begin{figure}
\centering
\includegraphics[totalheight=6cm,width=7cm]{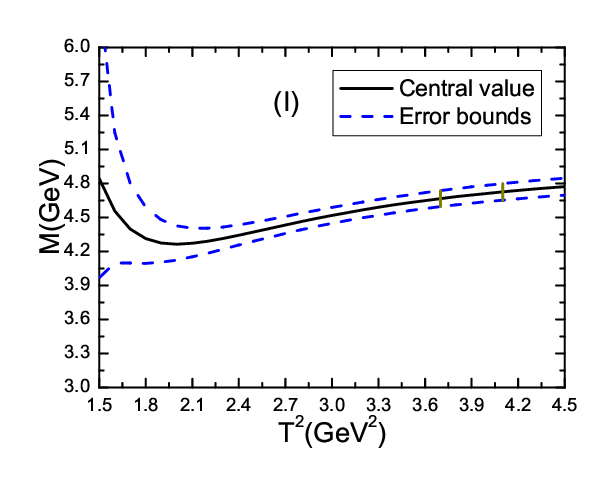}
\includegraphics[totalheight=6cm,width=7cm]{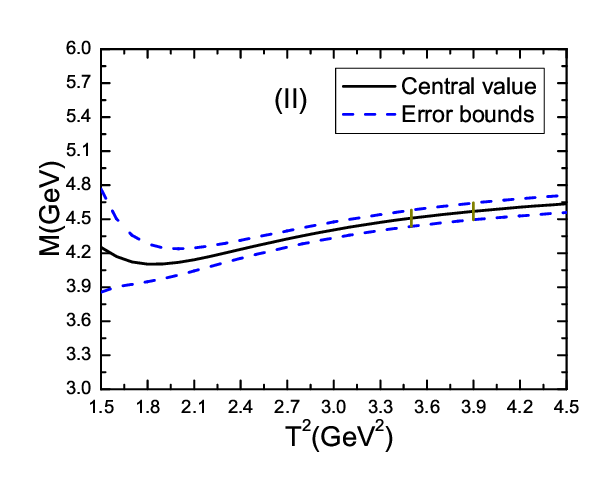}
\includegraphics[totalheight=6cm,width=7cm]{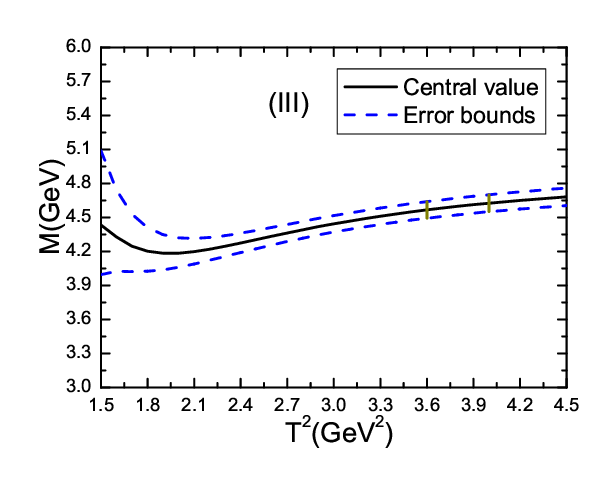}
\includegraphics[totalheight=6cm,width=7cm]{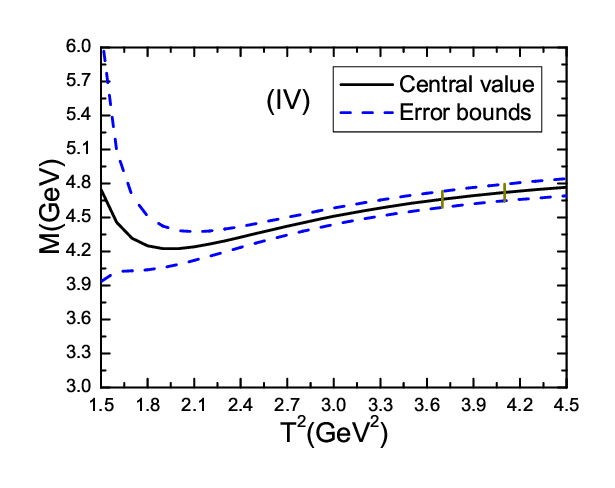}
\includegraphics[totalheight=6cm,width=7cm]{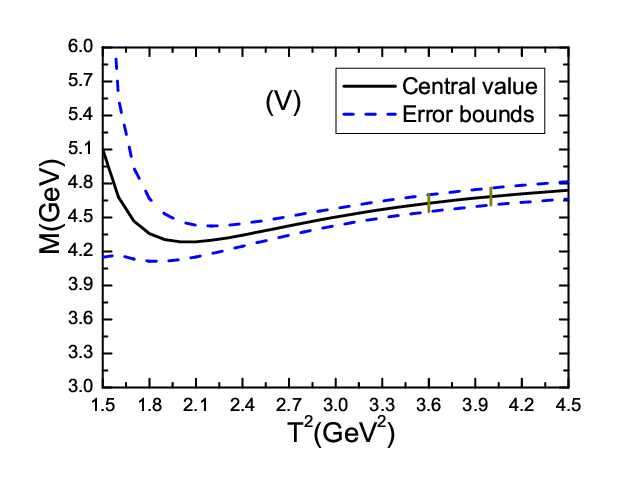}
  \caption{ The masses  with variations of the  Borel parameters $T^2$ for  the hidden-charm-doubly-strange pentaquark states, where the (I), (II), (III), (IV) and (V)  denote the
  $[qs][sc]\bar{c}$ ($0$, $1$, $1$, $\frac{5}{2}$), $[ss][qc]\bar{c}-[qs][sc]\bar{c}$ ($1$, $0$, $1$, $\frac{5}{2}$),  $[ss][qc]\bar{c}-[qs][sc]\bar{c}$ ($1$, $1$, $2$, $\frac{5}{2}$), $[qs][sc]\bar{c}$ ($1$, $0$, $1$, $\frac{5}{2}$) and $[ss][qc]\bar{c}$ ($1$, $1$, $2$, $\frac{5}{2}$)
    pentaquark states, respectively. }\label{mass-3-fig}
\end{figure}

Now we take  account of  all uncertainties  of the input   parameters,
and obtain  the masses and pole residues of
 the   hidden-charm-doubly-strange  pentaquark states with negative parity, which are shown explicitly in Figs.\ref{mass-1-fig}-\ref{mass-3-fig} and Table \ref{mass-Pcs}. From Tables \ref{Borel}-\ref{mass-Pcs}, we can obtain the conclusion  that the modified energy scale formula
 $\mu =\sqrt{M^2_{P}-(2{\mathbb{M}}_c)^2}-2{\mathbb{M}}_s$ is satisfied very well.
 The energy scale formula can enhance the pole contributions remarkably  and improve the convergent behavior of the operator product expansion  remarkably \cite{mole-penta-11,WangZG-Review}, which is the unique feature of our works. Direct comparisons indicate that without resorting to the energy scale formula, we could only obtain  bad convergent behavior of the operator product expansion or  poor pole contributions for the multiquark states \cite{mole-penta-11}.

In Figs.\ref{mass-1-fig}-\ref{mass-3-fig}, we plot the masses of the hidden-charm-doubly-strange pentaquark states with the isospin $I=\frac{1}{2}$ via variations of the Borel parameters, where the regions between the two short perpendicular lines are the Borel windows. In the Borel windows, the uncertainties $\frac{\delta M}{M}< 1\%$, there appear flat platforms indeed.  At the present time, there is no experimental candidate for those $P_{css}$ states to be compared to in those figures.

\begin{figure}
\centering
\includegraphics[totalheight=7cm,width=10cm]{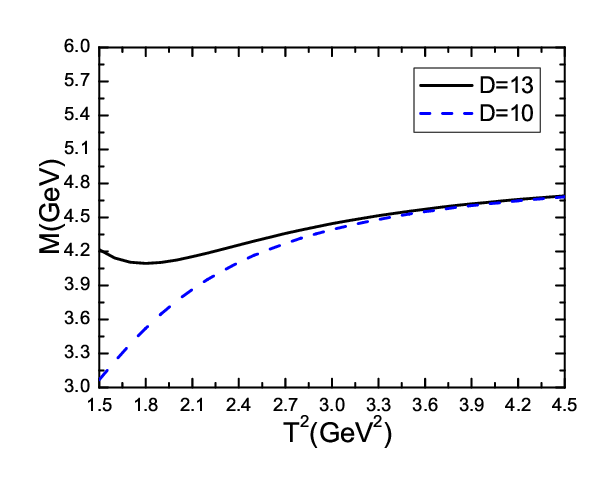}
  \caption{ The mass  with variation of the  Borel parameter $T^2$ for  the hidden-charm pentaquark state $[qs][sc]\bar{c}$ ($0$, $0$, $0$, $\frac{1}{2}$), where the $D=10$ and $13$ denote truncations of the operator product expansion, and the central values of other input parameters are implied. }\label{mass-D10-fig}
\end{figure}

In Fig.\ref{mass-D10-fig}, we plot the mass of the hidden-charm pentaquark state $[qs][sc]\bar{c}$ ($0$, $0$, $0$, $\frac{1}{2}$) with   variation of the  Borel parameter $T^2$ for truncations  of the operator product expansion up to the  vacuum condensates of dimension $10$ and $13$, respectively.   From the figure, we can see explicitly that
the vacuum condensates of dimension $11$ and $13$ play an important role in obtaining the flat platform, and we should update the old calculations \cite{Wang1508-EPJC,WangHuang-EPJC-1508-12,WangZG-EPJC-1509-12,
 WangZG-NPB-1512-32,WangZhang-APPB}.

From Table \ref{mass-Pcs}, we can see explicitly that the uncertainties of the predicted masses $\delta M= 0.10\,\rm{GeV}$ or $0.11\,\rm{GeV}$, which correspond  to the  uncertainties of the continuum threshold parameters $\delta \sqrt{s_0}=0.10\,\rm{GeV}$, the central values, upper bounds and lower bounds have one-to-one correspondences,
\begin{eqnarray}
\sqrt{s_0}-M&\approx&\left(\sqrt{s_0}+\delta \sqrt{s_0}\right)-\left(M+\delta M\right)\approx\left(\sqrt{s_0}-\delta \sqrt{s_0}\right)-\left(M-\delta M\right)\, .
\end{eqnarray}
A slightly larger uncertainties $\delta \sqrt{s_0}$ would destroy such a relation for the energy gaps between the ground states and first radial excited states. All in all, the present calculations are self-consistent.

From Table \ref{mass-Pcs}, also in our previous works \cite{WangZG-Pc12-Jpsip,WangZG-Pc12-JpsiLambda},  we could  obtain the conclusion confidently that the lowest hidden-charm pentaquark states are not of the $SS\bar{c}$ type, but are of the $AA\bar{c}$ type, it is not suitable to refer to the scalar and axialvector diquarks as the "good" and "bad" diquarks, respectively. And the scalar diquarks are not more stable than the axialvector diquarks. If we choose the same valence quarks and adopt the theoretical framework of the QCD sum rules, the $qq^\prime$-type scalar diquarks have slightly larger masses than the corresponding  axialvector diquarks \cite{WangZG-L-diquark-CTP}, while the $qQ$-type scalar and axialvector diquarks have almost degenerated masses \cite{WangZG-HL-diquark-EPJC,ZhangAL-HL-diquark-PRD}. We should bear in mind that the physical hadron masses are not simple summaries of the masses of all the constituents, but result from complex dynamics of the full QCD. The axialvector diquarks also serve as stable configurations in building the hidden-charm pentaquark states.

We can take the pole residues as elementary  input parameters and study the two-body strong decays,
 \begin{eqnarray}
P_{css}&\to& \bar{D}_s\Xi_c(4437/4440)\, , \,\bar{D}^*_s\Xi_c(4580/4583)\, , \, \bar{D}\Omega_c(4560/4564)\, , \, \bar{D}^*\Omega_c(4702/4705)\, ,  \nonumber\\
&&\,J/\psi \Xi(4412/4419) \, , \, \eta_c \Xi(4299/4306) \, ,
\end{eqnarray}
 with the three-point QCD sum rules to estimate the decay widths and select the
optimal channels to search for those pentaquark states $P_{css}$, where we present the thresholds of the meson-baryon pairs in the bracket and with the unit $\rm{MeV}$. As the two octets ${\mathbf{8}}_1$ and ${\mathbf{8}}_2$ in Eq.\eqref{two-octet} could mix with each other, the decays to the charmonium plus octet baryon are allowed.
Naively, we expect to observe the $P_c$, $P_{cs}$ and $P_{css}$ states in the weak decays of the ground state bottom baryons,
\begin{eqnarray}
\Lambda_b^0&\to& P_c^+K^- \to J/\psi p \,K^-\, ,\nonumber\\
&\to& P_{cs}^0\phi \to J/\psi \Lambda^0\, \phi \, ,\nonumber\\
\Xi_b^-&\to& P_{cs}^0 K^- \to J/\psi \Lambda^0\, K^-\, ,\nonumber\\
&\to&P_{css}^-\phi \to J/\psi \Xi^-  \phi \, ,
\end{eqnarray}
through the CKM favored weak process $b \to c\bar{c}s$ at the quark level with the effective Hamiltonian $H_w$,
\begin{eqnarray}
H_w &=& \frac{G_F}{\sqrt{2}}\,V_{c b} V_{c s}^*
\Big[ C_1(\mu)\, \overline{c}_{i}\Gamma_\mu b_{j}
\,\overline{s}_{j}\Gamma^\mu c_{i} +  C_2(\mu)\, \overline{c}_{i}\Gamma_\mu b_{i}
\,\overline{s}_{j}\Gamma^\mu c_{j}\Big]   \, ,\nonumber\\
&=& \frac{G_F}{\sqrt{2}}\,V_{c b} V_{c s}^*
\Big[ \widetilde{C}_1(\mu)\, \overline{c}\Gamma_\mu c\,
\overline{s}\Gamma^\mu b +  \widetilde{C}_2(\mu)\, \overline{c}\Gamma_\mu\frac{\lambda^a}{2} c\,
\overline{s}\Gamma^\mu\frac{\lambda^a}{2} b\Big]   \, ,
\end{eqnarray}
where $\Gamma_\mu=\gamma_\mu(1-\gamma_5)$, the $G_F$ is the Fermi coupling constant, the $V_{cb/cs}$ are the CKM matrix elements, the $i$ and $j$ are the color indexes, the $C_{1/2}$ ($\widetilde{C}_{1/2}$) are the
Wilson coefficients calculated at the renormalization scale $\mu
\sim O(m_b)$ \cite{AJBuras-1996}.
In fact, the $P_c(4312)$, $P_c(4337)$, $P_c(4380)$, $P_c(4440)$ and $P_c(4457)$  were observed in the $J/\psi p$ invariant mass spectrum, while the $P_{cs}(4338)$ and $P_{cs}(4459)$ were observed in the $J/\psi \Lambda$ invariant mass spectrum. We expect that the $P_{css}$ states could be observed in the $J/\psi \Xi$ invariant mass spectrum in the $\Xi_b$ decays.

\section{Conclusion}
 In the present work, we construct the color $\bar{\mathbf{3}}\bar{\mathbf{3}}\bar{\mathbf{3}}$ type local five-quark currents with the light quarks $qss$ in the flavor octet, and  investigate  the $qssc\bar{c}$ pentaquark states via  the QCD sum rules in a comprehensive way, and we emphasize that we achieve two light-flavor octets, which could mix with each other. We take account of   the    vacuum condensates up to dimension $13$ in a consistent way in our unique scheme, obtain the QCD spectral densities and select the contributions from the
 hidden-charm pentaquark states $P_{css}$ with the negative parity without contaminations,
 then resort to  the modified energy scale formula $\mu=\sqrt{M_{P}-(2{\mathbb{M}}_c)^2}-2{\mathbb{M}}_s$ to choose  the best  energy scales of the QCD spectral densities. At last, we obtain the mass spectrum of the hidden-charm-doubly-strange pentaquark states with the quantum numbers $IJ^{P}=\frac{1}{2}{\frac{1}{2}}^-$, $\frac{1}{2}{\frac{3}{2}}^-$, $\frac{1}{2}{\frac{5}{2}}^-$, which can be confronted to the experimental data in the future, especially in the process $\Xi_b^- \to P_{css}^-\phi \to J/\psi \Xi^-  \phi $, to examine the nature of those $P$ states. As a byproduct, we obtain the conclusion  that the lowest pentaquark states are not of the scalar-diquark-scalar-diquark-antiquark type, it is not suitable to refer to  the scalar and axialvector diquarks as the "good" and "bad" diquarks, respectively.

\section*{Acknowledgements}
This  work is supported by National Natural Science Foundation, Grant Number  12575083.

\end{document}